\documentclass[twocolumn]{IEEEtran-v17}
	
\pdfoutput=1 

\newlength{\figwidth}
\usepackage{xspace}
\usepackage[nosort]{cite}        
\usepackage{url} 
\usepackage[intlimits]{amsmath}
\usepackage{bbm}
\usepackage{graphicx}
\usepackage{paralist}
\usepackage{fancyref}
\usepackage[stretch=16,shrink=16,step=4]{microtype}

\makeatletter
\def\@IEEEinterspaceratioM{0.265}
\def\@IEEEinterspaceMINratioM{0.1651}
\def\@IEEEinterspaceMAXratioM{0.38}

\def\@IEEEinterspaceratioB{0.31}
\def\@IEEEinterspaceMINratioB{0.19}
\def\@IEEEinterspaceMAXratioB{0.38}
\@IEEEtunefonts
\makeatother

\usepackage{amssymb}
\usepackage{amsfonts}
\usepackage{mathrsfs}
\usepackage{xspace}
\usepackage{bm}
\usepackage{upgreek}

\newcommand{\safemath}[2]{\newcommand{#1}{\ensuremath{#2}\xspace}}

\safemath{\bma}{\mathbf{a}}
\safemath{\bmb}{\mathbf{b}}
\safemath{\bmc}{\mathbf{c}}
\safemath{\bmd}{\mathbf{d}}
\safemath{\bme}{\mathbf{e}}
\safemath{\bmf}{\mathbf{f}}
\safemath{\bmg}{\mathbf{g}}
\safemath{\bmh}{\mathbf{h}}
\safemath{\bmi}{\mathbf{i}}
\safemath{\bmj}{\mathbf{j}}
\safemath{\bmk}{\mathbf{k}}
\safemath{\bml}{\mathbf{l}}
\safemath{\bmm}{\mathbf{m}}
\safemath{\bmn}{\mathbf{n}}
\safemath{\bmo}{\mathbf{o}}
\safemath{\bmp}{\mathbf{p}}
\safemath{\bmq}{\mathbf{q}}
\safemath{\bmr}{\mathbf{r}}
\safemath{\bms}{\mathbf{s}}
\safemath{\bmt}{\mathbf{t}}
\safemath{\bmu}{\mathbf{u}}
\safemath{\bmv}{\mathbf{v}}
\safemath{\bmw}{\mathbf{w}}
\safemath{\bmx}{\mathbf{x}}
\safemath{\bmy}{\mathbf{y}}
\safemath{\bmz}{\mathbf{z}}
\safemath{\bmzero}{\mathbf{0}}
\safemath{\bmone}{\mathbf{1}}

\bmdefine{\biad}{a}
\bmdefine{\bibd}{b}
\bmdefine{\bicd}{c}
\bmdefine{\bidd}{d}
\bmdefine{\bied}{e}
\bmdefine{\bifd}{f}
\bmdefine{\bigd}{g}
\bmdefine{\bihd}{h}
\bmdefine{\biid}{i}
\bmdefine{\bijd}{j}
\bmdefine{\bikd}{k}
\bmdefine{\bild}{l}
\bmdefine{\bimd}{m}
\bmdefine{\bind}{n}
\bmdefine{\biod}{o}
\bmdefine{\bipd}{p}
\bmdefine{\biqd}{q}
\bmdefine{\bird}{r}
\bmdefine{\bisd}{s}
\bmdefine{\bitd}{t}
\bmdefine{\biud}{u}
\bmdefine{\bivd}{v}
\bmdefine{\biwd}{w}
\bmdefine{\bixd}{x}
\bmdefine{\biyd}{y}
\bmdefine{\bizd}{z}

\bmdefine{\bixid}{\xi}
\bmdefine{\bilambdad}{\lambda}
\bmdefine{\bimud}{\mu}
\bmdefine{\bithetad}{\theta}
\bmdefine{\biphid}{\phi}
\bmdefine{\bialphad}{\alpha}
\bmdefine{\bibetad}{\beta}
\bmdefine{\bisigma}{\sigma}

\safemath{\bmia}{\biad}
\safemath{\bmib}{\bibd}
\safemath{\bmic}{\bicd}
\safemath{\bmid}{\bidd}
\safemath{\bmie}{\bied}
\safemath{\bmif}{\bifd}
\safemath{\bmig}{\bigd}
\safemath{\bmih}{\bihd}
\safemath{\bmii}{\biid}
\safemath{\bmij}{\bijd}
\safemath{\bmik}{\bikd}
\safemath{\bmil}{\bild}
\safemath{\bmim}{\bimd}
\safemath{\bmin}{\bind}
\safemath{\bmio}{\biod}
\safemath{\bmip}{\bipd}
\safemath{\bmiq}{\biqd}
\safemath{\bmir}{\bird}
\safemath{\bmis}{\bisd}
\safemath{\bmit}{\bitd}
\safemath{\bmiu}{\biud}
\safemath{\bmiv}{\bivd}
\safemath{\bmiw}{\biwd}
\safemath{\bmix}{\bixd}
\safemath{\bmiy}{\biyd}
\safemath{\bmiz}{\bizd}

\safemath{\bmalpha}{\bialphad}
\safemath{\bmbeta}{\bibetad}
\safemath{\bmxi}{\bixid}
\safemath{\bmlambda}{\bilambdad}
\safemath{\bmmu}{\bimud}
\safemath{\bmtheta}{\bithetad}
\safemath{\bmphi}{\biphid}
\safemath{\bmsigma}{\bisigma}

\safemath{\bA}{\mathbf{A}}
\safemath{\bB}{\mathbf{B}}
\safemath{\bC}{\mathbf{C}}
\safemath{\bD}{\mathbf{D}}
\safemath{\bE}{\mathbf{E}}
\safemath{\bF}{\mathbf{F}}
\safemath{\bG}{\mathbf{G}}
\safemath{\bH}{\mathbf{H}}
\safemath{\bI}{\mathbf{I}}
\safemath{\bJ}{\mathbf{J}}
\safemath{\bK}{\mathbf{K}}
\safemath{\bL}{\mathbf{L}}
\safemath{\bM}{\mathbf{M}}
\safemath{\bN}{\mathbf{N}}
\safemath{\bO}{\mathbf{O}}
\safemath{\bP}{\mathbf{P}}
\safemath{\bQ}{\mathbf{Q}}
\safemath{\bR}{\mathbf{R}}
\safemath{\bS}{\mathbf{S}}
\safemath{\bT}{\mathbf{T}}
\safemath{\bU}{\mathbf{U}}
\safemath{\bV}{\mathbf{V}}
\safemath{\bW}{\mathbf{W}}
\safemath{\bX}{\mathbf{X}}
\safemath{\bY}{\mathbf{Y}}
\safemath{\bZ}{\mathbf{Z}}

\safemath{\bZero}{\mathbf{0}}
\safemath{\bOne}{\mathbf{1}}
\safemath{\bDelta}{\mathbf{\Delta}}
\safemath{\bLambda}{\mathbf{\Lambda}}
\safemath{\bPhi}{\mathbf{\Phi}}
\safemath{\bSigma}{\mathbf{\Sigma}}
\safemath{\bOmega}{\mathbf{\Omega}}
\safemath{\bTheta}{\mathbf{\Theta}}

\bmdefine{\biAd}{A}
\bmdefine{\biBd}{B}
\bmdefine{\biCd}{C}
\bmdefine{\biDd}{D}
\bmdefine{\biEd}{E}
\bmdefine{\biFd}{F}
\bmdefine{\biGd}{G}
\bmdefine{\biHd}{H}
\bmdefine{\biId}{I}
\bmdefine{\biJd}{J}
\bmdefine{\biKd}{K}
\bmdefine{\biLd}{L}
\bmdefine{\biMd}{M}
\bmdefine{\biOd}{N}
\bmdefine{\biPd}{O}
\bmdefine{\biQd}{P}
\bmdefine{\biRd}{R}
\bmdefine{\biSd}{S}
\bmdefine{\biTd}{T}
\bmdefine{\biUd}{U}
\bmdefine{\biVd}{V}
\bmdefine{\biWd}{W}
\bmdefine{\biXd}{X}
\bmdefine{\biYd}{Y}
\bmdefine{\biZd}{Z}

\bmdefine{\biDelta}{\Delta}
\bmdefine{\biLambda}{\Lambda}
\bmdefine{\biPhi}{\Phi}
\bmdefine{\biSigma}{\Sigma}
\bmdefine{\biOmega}{\Omega}
\bmdefine{\biTheta}{\Theta}

\safemath{\bimA}{\biAd}
\safemath{\bimB}{\biBd}
\safemath{\bimC}{\biCd}
\safemath{\bimD}{\biDd}
\safemath{\bimE}{\biEd}
\safemath{\bimF}{\biFd}
\safemath{\bimG}{\biGd}
\safemath{\bimH}{\biHd}
\safemath{\bimI}{\biId}
\safemath{\bimJ}{\biJd}
\safemath{\bimK}{\biKd}
\safemath{\bimL}{\biLd}
\safemath{\bimM}{\biMd}
\safemath{\bimN}{\biNd}
\safemath{\bimO}{\biOd}
\safemath{\bimP}{\biPd}
\safemath{\bimQ}{\biQd}
\safemath{\bimR}{\biRd}
\safemath{\bimS}{\biSd}
\safemath{\bimT}{\biTd}
\safemath{\bimU}{\biUd}
\safemath{\bimV}{\biVd}
\safemath{\bimW}{\biWd}
\safemath{\bimX}{\biXd}
\safemath{\bimY}{\biYd}
\safemath{\bimZ}{\biZd}

\safemath{\bimDelta}{\biDelta}
\safemath{\bimLambda}{\biLambda}
\safemath{\bimPhi}{\biPhi}
\safemath{\bimSigma}{\biSigma}
\safemath{\bimOmega}{\biOmega}
\safemath{\bimTheta}{\biTheta}

\safemath{\setA}{\mathcal{A}}
\safemath{\setB}{\mathcal{B}}
\safemath{\setC}{\mathcal{C}}
\safemath{\setD}{\mathcal{D}}
\safemath{\setE}{\mathcal{E}}
\safemath{\setF}{\mathcal{F}}
\safemath{\setG}{\mathcal{G}}
\safemath{\setH}{\mathcal{H}}
\safemath{\setI}{\mathcal{I}}
\safemath{\setJ}{\mathcal{J}}
\safemath{\setK}{\mathcal{K}}
\safemath{\setL}{\mathcal{L}}
\safemath{\setM}{\mathcal{M}}
\safemath{\setN}{\mathcal{N}}
\safemath{\setO}{\mathcal{O}}
\safemath{\setP}{\mathcal{P}}
\safemath{\setQ}{\mathcal{Q}}
\safemath{\setR}{\mathcal{R}}
\safemath{\setS}{\mathcal{S}}
\safemath{\setT}{\mathcal{T}}
\safemath{\setU}{\mathcal{U}}
\safemath{\setV}{\mathcal{V}}
\safemath{\setW}{\mathcal{W}}
\safemath{\setX}{\mathcal{X}}
\safemath{\setY}{\mathcal{Y}}
\safemath{\setZ}{\mathcal{Z}}
\safemath{\emptySet}{\varnothing}

\safemath{\colA}{\mathscr{A}}
\safemath{\colB}{\mathscr{B}}
\safemath{\colC}{\mathscr{C}}
\safemath{\colD}{\mathscr{D}}
\safemath{\colE}{\mathscr{E}}
\safemath{\colF}{\mathscr{F}}
\safemath{\colG}{\mathscr{G}}
\safemath{\colH}{\mathscr{H}}
\safemath{\colI}{\mathscr{I}}
\safemath{\colJ}{\mathscr{J}}
\safemath{\colK}{\mathscr{K}}
\safemath{\colL}{\mathscr{L}}
\safemath{\colM}{\mathscr{M}}
\safemath{\colN}{\mathscr{N}}
\safemath{\colO}{\mathscr{O}}
\safemath{\colP}{\mathscr{P}}
\safemath{\colQ}{\mathscr{Q}}
\safemath{\colR}{\mathscr{R}}
\safemath{\colS}{\mathscr{S}}
\safemath{\colT}{\mathscr{T}}
\safemath{\colU}{\mathscr{U}}
\safemath{\colV}{\mathscr{V}}
\safemath{\colW}{\mathscr{W}}
\safemath{\colX}{\mathscr{X}}
\safemath{\colY}{\mathscr{Y}}
\safemath{\colZ}{\mathscr{Z}}

\safemath{\opA}{\mathbb{A}}
\safemath{\opB}{\mathbb{B}}
\safemath{\opC}{\mathbb{C}}
\safemath{\opD}{\mathbb{D}}
\safemath{\opE}{\mathbb{E}}
\safemath{\opF}{\mathbb{F}}
\safemath{\opG}{\mathbb{G}}
\safemath{\opH}{\mathbb{H}}
\safemath{\opI}{\mathbb{I}}
\safemath{\opJ}{\mathbb{J}}
\safemath{\opK}{\mathbb{K}}
\safemath{\opL}{\mathbb{L}}
\safemath{\opM}{\mathbb{M}}
\safemath{\opN}{\mathbb{N}}
\safemath{\opO}{\mathbb{O}}
\safemath{\opP}{\mathbb{P}}
\safemath{\opQ}{\mathbb{Q}}
\safemath{\opR}{\mathbb{R}}
\safemath{\opS}{\mathbb{S}}
\safemath{\opT}{\mathbb{T}}
\safemath{\opU}{\mathbb{U}}
\safemath{\opV}{\mathbb{V}}
\safemath{\opW}{\mathbb{W}}
\safemath{\opX}{\mathbb{X}}
\safemath{\opY}{\mathbb{Y}}
\safemath{\opZ}{\mathbb{Z}}
\safemath{\opZero}{\mathbb{O}}
\safemath{\identityop}{\opI}

\safemath{\veca}{\bma}
\safemath{\vecb}{\bmb}
\safemath{\vecc}{\bmc}
\safemath{\vecd}{\bmd}
\safemath{\vece}{\bme}
\safemath{\vecf}{\bmf}
\safemath{\vecg}{\bmg}
\safemath{\vech}{\bmh}
\safemath{\veci}{\bmi}
\safemath{\vecj}{\bmj}
\safemath{\veck}{\bmk}
\safemath{\vecl}{\bml}
\safemath{\vecm}{\bmm}
\safemath{\vecn}{\bmn}
\safemath{\veco}{\bmo}
\safemath{\vecp}{\bmmp}
\safemath{\vecq}{\bmq}
\safemath{\vecr}{\bmr}
\safemath{\vecs}{\bms}
\safemath{\vect}{\bmt}
\safemath{\vecu}{\bmu}
\safemath{\vecv}{\bmv}
\safemath{\vecw}{\bmw}
\safemath{\vecx}{\bmx}
\safemath{\vecy}{\bmy}
\safemath{\vecz}{\bmz}

\safemath{\veczero}{\bmzero}
\safemath{\vecone}{\bmone}
\safemath{\vecalpha}{\bmalpha}
\safemath{\vecbeta}{\bmbeta}
\safemath{\vecxi}{\bmxi}
\safemath{\veclambda}{\bmlambda}
\safemath{\vecmu}{\bmmu}
\safemath{\vectheta}{\bmtheta}
\safemath{\vecphi}{\bmphi}
\safemath{\vecsigma}{\bmsigma}

\safemath{\matA}{\bA}
\safemath{\matB}{\bB}
\safemath{\matC}{\bC}
\safemath{\matD}{\bD}
\safemath{\matE}{\bE}
\safemath{\matF}{\bF}
\safemath{\matG}{\bG}
\safemath{\matH}{\bH}
\safemath{\matI}{\bI}
\safemath{\matJ}{\bJ}
\safemath{\matK}{\bK}
\safemath{\matL}{\bL}
\safemath{\matM}{\bM}
\safemath{\matN}{\bN}
\safemath{\matO}{\bO}
\safemath{\matP}{\bP}
\safemath{\matQ}{\bQ}
\safemath{\matR}{\bR}
\safemath{\matS}{\bS}
\safemath{\matT}{\bT}
\safemath{\matU}{\bU}
\safemath{\matV}{\bV}
\safemath{\matW}{\bW}
\safemath{\matX}{\bX}
\safemath{\matY}{\bY}
\safemath{\matZ}{\bZ}
\safemath{\matzero}{\bmzero}

\safemath{\matDelta}{\bDelta}
\safemath{\matLambda}{\bLambda}
\safemath{\matPhi}{\bPhi}
\safemath{\matSigma}{\bSigma}
\safemath{\matOmega}{\bOmega}
\safemath{\matTheta}{\bTheta}

\safemath{\matidentity}{\matI}
\safemath{\matone}{\matO}

\safemath{\rnda}{A}
\safemath{\rndb}{B}
\safemath{\rndc}{C}
\safemath{\rndd}{D}
\safemath{\rnde}{E}
\safemath{\rndf}{F}
\safemath{\rndg}{G}
\safemath{\rndh}{H}
\safemath{\rndi}{I}
\safemath{\rndj}{J}
\safemath{\rndk}{K}
\safemath{\rndl}{L}
\safemath{\rndm}{M}
\safemath{\rndn}{N}
\safemath{\rndo}{O}
\safemath{\rndp}{P}
\safemath{\rndq}{Q}
\safemath{\rndr}{R}
\safemath{\rnds}{S}
\safemath{\rndt}{T}
\safemath{\rndu}{U}
\safemath{\rndv}{V}
\safemath{\rndw}{W}
\safemath{\rndx}{X}
\safemath{\rndy}{Y}
\safemath{\rndz}{Z}

\safemath{\rveca}{\bimA}
\safemath{\rvecb}{\bimB}
\safemath{\rvecc}{\bimC}
\safemath{\rvecd}{\bimD}
\safemath{\rvece}{\bimE}
\safemath{\rvecf}{\bimF}
\safemath{\rvecg}{\bimG}
\safemath{\rvech}{\bimH}
\safemath{\rveci}{\bimI}
\safemath{\rvecj}{\bimJ}
\safemath{\rveck}{\bimK}
\safemath{\rvecl}{\bimL}
\safemath{\rvecm}{\bimM}
\safemath{\rvecn}{\bimN}
\safemath{\rveco}{\bomO}
\safemath{\rvecp}{\bimP}
\safemath{\rvecq}{\bimQ}
\safemath{\rvecr}{\bimR}
\safemath{\rvecs}{\bimS}
\safemath{\rvect}{\bimT}
\safemath{\rvecu}{\bimU}
\safemath{\rvecv}{\bimV}
\safemath{\rvecw}{\bimW}
\safemath{\rvecx}{\bimX}
\safemath{\rvecy}{\bimY}
\safemath{\rvecz}{\bimZ}

\safemath{\rvecalpha}{\bmalpha}
\safemath{\rvecbeta}{\bmbeta}
\safemath{\rvecxi}{\bmxi}
\safemath{\rveclambda}{\bmlambda}
\safemath{\rvecmu}{\bmmu}
\safemath{\rvectheta}{\bmtheta}
\safemath{\rvecphi}{\bmphi}

\safemath{\rmatA}{\bimA}
\safemath{\rmatB}{\bimB}
\safemath{\rmatC}{\bimC}
\safemath{\rmatD}{\bimD}
\safemath{\rmatE}{\bimE}
\safemath{\rmatF}{\bimF}
\safemath{\rmatG}{\bimG}
\safemath{\rmatH}{\bimH}
\safemath{\rmatI}{\bimI}
\safemath{\rmatJ}{\bimJ}
\safemath{\rmatK}{\bimK}
\safemath{\rmatL}{\bimL}
\safemath{\rmatM}{\bimM}
\safemath{\rmatN}{\bimN}
\safemath{\rmatO}{\bimO}
\safemath{\rmatP}{\bimP}
\safemath{\rmatQ}{\bimQ}
\safemath{\rmatR}{\bimR}
\safemath{\rmatS}{\bimS}
\safemath{\rmatT}{\bimT}
\safemath{\rmatU}{\bimU}
\safemath{\rmatV}{\bimV}
\safemath{\rmatW}{\bimW}
\safemath{\rmatX}{\bimX}
\safemath{\rmatY}{\bimY}
\safemath{\rmatZ}{\bimZ}

\safemath{\rmatDelta}{\bimDelta}
\safemath{\rmatLambda}{\bimLambda}
\safemath{\rmatPhi}{\bimPhi}
\safemath{\rmatSigma}{\bimSigma}
\safemath{\rmatOmega}{\bimOmega}
\safemath{\rmatTheta}{\bimTheta}

\usepackage{amssymb}
\usepackage{amsfonts}
\usepackage{mathrsfs}
\usepackage{xspace}
\usepackage{bm}
\usepackage{textcomp}

\newenvironment{textbmatrix}{	\setlength{\arraycolsep}{2.5pt}%
								\big[\begin{matrix}}{\end{matrix}\big]%
								\raisebox{0.08ex}{\vphantom{M}}}

\def\be{\begin{equation}}
\def\ee{\end{equation}}
\def\een{\nonumber \end{equation}}
\def\mat{\begin{bmatrix}}
\def\emat{\end{bmatrix}}
\def\btm{\begin{textbmatrix}}
\def\etm{\end{textbmatrix}}

\def\ba#1\ea{\begin{align}#1\end{align}}
\def\bas#1\eas{\begin{align*}#1\end{align*}}
\def\bs#1\es{\begin{split}#1\end{split}}
\def\bml#1\eml{\begin{multline}#1\end{multline}}
\def\bg#1\eg{\begin{gather}#1\end{gather}} 
\def\bi#1\ei{\begin{itemize}#1\end{itemize}}

\newcommand{\lefto}{\mathopen{}\left}

\DeclareMathOperator{\tr}{tr}				% trace
\DeclareMathOperator{\dg}{\opD}				% diagonal matrix
\DeclareMathOperator{\rank}{rank}			% rank of a matrix
\DeclareMathOperator{\kron}{\otimes}			% Kroneker Product
\DeclareMathOperator{\had}{\odot}			% Hadamard Product
\DeclareMathOperator{\four}{\opF}			% Fourier transform
\DeclareMathOperator{\Exop}{\opE}			% expectation operator
\DeclareMathOperator{\landauo}{\mathit{o}}

\newcommand{\Ex}[2]{\ensuremath{\Exop_{#1}\lefto[#2\right]}} 	% expectation
\newcommand{\abs}[1]{\left\lvert#1\right\rvert}		% absolute value
\newcommand{\vecnorm}[1]{\lVert#1\rVert}		% vector norm
\newcommand{\conj}[1]{\ensuremath{#1^{*}}} 	% conjugate 		
\newcommand{\tp}[1]{\ensuremath{#1^{T}}} 		% transpose
\newcommand{\herm}[1]{\ensuremath{#1^{H}}} 	% hermitian transpose
\newcommand{\inv}[1]{\ensuremath{#1^{-1}}} 	% inverse
\newcommand{\sqm}[1]{\ensuremath{#1^{1/2}}}	% square-root matrix
\safemath{\dirac}{\delta}					% Dirac delta
\safemath{\krond}{\dirac}					% Kronecker delta
\newcommand{\allz}[2]{\ensuremath{#1=0,1,\ldots,#2-1}}% all definition by Moritz
\newcommand{\sumz}[2]{\ensuremath{\sum_{#1=0}^{#2-1}}}

\newcommand{\logdet}[1]{\log \det\lefto(#1\right)} % log det function
\safemath{\upto}{\uparrow}
\safemath{\downto}{\downarrow}
\safemath{\iu}{i}							% imaginary unit
\safemath{\maj}{\prec}

\safemath{\hilseqspace}{l^{2}}				% Hilbert sequence space
\newcommand{\banachfunspace}[1]{\setL^{#1}}	% Banach function space
\safemath{\hilfunspace}{\banachfunspace{2}}	% Hilbert function space
\safemath{\SNR}{\text{\sc snr}} 				% signal to noise ratio
\safemath{\No}{N_0}							% noise spectral density
\safemath{\Es}{E_s}							% energy per symbol
\safemath{\Eb}{E_b}							% energy per bit
\safemath{\EbNo}{\frac{\Eb}{\No}}
\safemath{\EsNo}{\frac{\Es}{\No}}

\DeclareMathOperator{\CHop}{\ensuremath{\opH}} % channel operator
\safemath{\tvir}{\rndh_{\CHop}}				% time-varying impulse response
\safemath{\tvtf}{\rndl_{\CHop}}				% 	-''- transfer function
\safemath{\spf}{\rnds_{\CHop}}				% spreading function
\safemath{\bff}{H_{\CHop}}					% bi-freuqency function

\safemath{\ircf}{r_{h}}						% impulse response correlation fn.
\safemath{\tftvcf}{r_{s}}					% scattering function
\safemath{\tfcf}{r_{l}}						% time-frequency correlation fn.
\safemath{\bfcf}{r_{H}}						% bi-frequency correlation fn.

\safemath{\tcorr}{c_h}						% time-correlation function
\safemath{\scf}{c_{s}}						% spreading function
\safemath{\tfcorr}{c_{l}}					% transfer-function correlation
\safemath{\fcorr}{c_{H}}						% frequency-correlation function

\safemath{\mi}{I}							% mutual information
\safemath{\capacity}{C}						% capacity

\newcommand{\iid}{i.i.d.\@\xspace}

\safemath{\normal}{\mathcal{N}}			% normal distribution
\safemath{\jpg}{\mathcal{CN}}			% jointly proper Gaussian
\safemath{\mchain}{\leftrightarrow}		% Markov chain
\newcommand{\given}{\,\vert\,}				% conditioning
\safemath{\dB}{\,\mathrm{dB}}
\safemath{\dBm}{\,\mathrm{dBm}}
\safemath{\Hz}{\,\mathrm{Hz}}
\safemath{\kHz}{\,\mathrm{kHz}}
\safemath{\MHz}{\,\mathrm{MHz}}
\safemath{\GHz}{\,\mathrm{GHz}}
\safemath{\s}{\,\mathrm{s}}
\safemath{\ms}{\,\mathrm{ms}}
\safemath{\mus}{\,\mathrm{\text{\textmu}s}}
\safemath{\ns}{\,\mathrm{ns}}
\safemath{\meter}{\,\mathrm{m}}
\safemath{\mm}{\,\mathrm{mm}}
\safemath{\cm}{\,\mathrm{cm}}
\safemath{\m}{\,\mathrm{m}}
\safemath{\W}{\,\mathrm{W}}
\safemath{\mW}{\, \mathrm{mW}}
\safemath{\J}{\,\mathrm{J}}
\safemath{\K}{\,\mathrm{K}}
\safemath{\bit}{\,\mathrm{bit}}

\safemath{\define}{=}			% definition

\safemath{\equivalent}{\sim}
\safemath{\distas}{\sim}					% distributed according to
\safemath{\sdiff}{\Delta}				% symmetric set difference

\safemath{\reals}{\mathbb{R}}
\safemath{\positivereals}{\reals_{+}}
\safemath{\integers}{\mathbb{Z}}
\safemath{\posint}{\integers_{+}}
\safemath{\naturals}{\mathbb{N}}
\safemath{\posnaturals}{\naturals_{+}}
\safemath{\complexset}{\mathbb{C}}
\safemath{\rationals}{\mathbb{Q}}

\newcommand*{\fancyrefapplabelprefix}{app}		% Appendix
\newcommand*{\fancyrefthmlabelprefix}{thm}		% Theorem
\newcommand*{\fancyreflemlabelprefix}{lem}		% Lemma
\newcommand*{\fancyrefcorlabelprefix}{cor}		% Corolary
\newcommand*{\fancyrefdeflabelprefix}{def}		% Definition

\frefformat{vario}{\fancyrefseclabelprefix}{Section~#1}
\frefformat{vario}{\fancyrefthmlabelprefix}{Theorem~#1}
\frefformat{vario}{\fancyreflemlabelprefix}{Lemma~#1}
\frefformat{vario}{\fancyrefcorlabelprefix}{Corollary~#1}
\frefformat{vario}{\fancyrefdeflabelprefix}{Definition~#1}
\frefformat{vario}{\fancyreffiglabelprefix}{Fig.~#1}
\frefformat{vario}{\fancyrefapplabelprefix}{Appendix~#1}
\frefformat{vario}{\fancyrefeqlabelprefix}{(#1)}

\newtheorem{thm}{Theorem}
\newtheorem{lem}[thm]{Lemma}

\let\time\undefined
\let\tvir\undefined
\let\tvtf\undefined
\let\spf\undefined
\let\scf\undefined
\let\spread\undefined
\let\dg\undefined

\safemath{\time}{t}						% time variable
\safemath{\freq}{f}						% frequency variable
\safemath{\doppler}{\nu}					% Doppler variable
\safemath{\delay}{\tau}					% delay variable
\safemath{\dd}{\doppler,\delay}			% Doppler-delay shorthand
\safemath{\maxDoppler}{\nu_{0}}
\safemath{\maxDelay}{\tau_{0}}
\safemath{\dtime}{k}						% discrete time
\safemath{\dfreq}{n}						% discrete frequency
\safemath{\dtdf}{\dtime,\dfreq}
\safemath{\tstep}{T}						% grid spacing in time
\safemath{\fstep}{F}						% grid spacing in frequency
\safemath{\tfstep}{\tstep\fstep}
\safemath{\fslots}{N}					% number of tones
\safemath{\tslots}{K}					% number of time slots
\safemath{\tfslots}{\tslots\fslots}
\safemath{\bandwidth}{W}					% signal bandwidth 

\safemath{\ch}{\CHop}
\safemath{\ir}{h}						% channel impulse response
\safemath{\irp}{\ir[\dtdf]}
\safemath{\kernel}{k_{\CHop}}			% LTV kernel
\safemath{\kernelp}{\kernel(\time,\time')}
\safemath{\tvtf}{L_{\CHop}}				% time-varying transfer function
\safemath{\spf}{S_{\CHop}}				% spreading function
\safemath{\spfp}{\spf(\dd)}				% '' parametrized
\safemath{\scf}{C_{\CHop}}				% scattering function
\safemath{\scfp}{\scf(\dd)}				% ''	 parametrized
\safemath{\scfpsq}{\scf^{2}(\dd)}		% parameterized squared scattering 
\safemath{\spread}{\Delta_{\CHop}}		% total spread
\safemath{\eqspread}{\tilde{\Delta}}		% equivalent spread
\safemath{\peakiness}{\kappa_{\CHop}}    % peakiness measure
\safemath{\chcorr}{R}					% channel correlation function
\safemath{\chcorrp}{\chcorr[\dtdf]}
\safemath{\txcorr}{A}					% transmit correlation
\safemath{\txcorrmat}{\matA}				% transmit correlation matrix
\safemath{\unitarymat}{\matU}			% unitary matrix in sans serif
\safemath{\txcorrunit}{\unitarymat_{\!\txcorr}}% transmit correlation decomp.
\safemath{\txcorrspec}{\matSigma}		% transmit correlation spectrum
\safemath{\txev}{\sigma}					% transmit correlation eigenvalue
\safemath{\txevvec}{\vecsigma}			% transmit correlation eigenvalue vec
\safemath{\txevm}{\txev_{0}}				% max transmit correlation eigenvalue
\safemath{\rxcorr}{B}					% receive correlation
\safemath{\rxcorrmat}{\matB}				% receive correlation matrix
\safemath{\rxcorrunit}{\unitarymat_{\!\rxcorr}}	% receive correlation decomp.
\safemath{\rxcorrspec}{\matLambda}		% receive correlation spectrum
\safemath{\rxev}{\lambda}				% receive correlation eigenvalue
\safemath{\rxevp}{\lambda_{\rxindex}}		%	''	parameterized
\safemath{\rxevvec}{\veclambda}			% receive correlation eigenvalue vector
\safemath{\rxevm}{\rxev_{0}}				% max receive correlation eigenvalue
\safemath{\schcorrmat}{\covmat{}}		% SISO channel correlation matrix
\safemath{\chspecmat}{\matC}				% matrix-valued spectrum
\safemath{\chspecmatp}{\chspecmat(\specparam)}

\safemath{\tfchmat}{\bm{\mathsf{H}}}		% MIMO channel
\safemath{\tfchmatp}{\tfchmat[\dtdf]}		% 	`` parametrized
\safemath{\tfchmatw}{\bm{\mathsf{H}}_{\mathsf{w}}}	% white MIMO channel
\safemath{\tfchmatwp}{\tfchmatw[\dtdf]}	% 	'' parametrized
\safemath{\alttfchmatw}{\tilde{\tfchmat}_{\mathsf{w}}}	% alternate
\safemath{\alttfchmatwp}{\alttfchmatw[\dtdf]}	% 	'' parametrized
\safemath{\chvec}{\bmh}					% stacked MIMO channel vector
\safemath{\altchvec}{\tilde{\chvec}}     % alternative stacked MIMO vector
\safemath{\numtx}{M_{T}}					% number of transmit antennas
\safemath{\altnumtx}{Q}					% alt. number of transmit antennas	                                  
\safemath{\numrx}{M_{R}}					% number of receive antennas
\safemath{\txindex}{q}					% index tx antennas
\safemath{\rxindex}{r}					% index for rx antennas
\safemath{\mimo}{\rxindex,\txindex}		% mimo subscript
\safemath{\txdim}{\numtx\tslots\fslots}  % number of transmit dimensions
\safemath{\rxdim}{\numrx\tslots\fslots}  % number of receive dimensions
\safemath{\totdim}{\numtx\numrx\tslots\fslots}  % number of total dimensions
\safemath{\rxrank}{Q_{R}}				% rank of the receive correlation

\safemath{\logon}{g}						% Weyl-Heisenberg basis logon
\safemath{\logonp}{\logon_{\dtdf}(\time)}
\safemath{\inp}{x}						
\safemath{\inppct}{\inp(\time)}			% parametrized continuous-time input
\safemath{\inpp}{\inp[\dtdf]}			% parametrized discrete-time input
\safemath{\wgn}{w}						% white Gaussian noise
\safemath{\wgnp}{\wgn[\dtdf]}
\safemath{\outp}{y}						% channel output signal
\safemath{\outppct}{\outp(\time)}
\safemath{\outpp}{\outp[\dtdf]}
\safemath{\inpvec}{\vecx}				% channel input signal MIMO
\safemath{\altinpvec}{\tilde{\inpvec}}   % alternative input signal MIMO
\safemath{\inpmat}{\matX}				% input channel matrix MIMO
\safemath{\altinpmat}{\tilde{\inpmat}}   % alternative input channel matrix MIMO
\safemath{\outpvec}{\vecy}				% output vector MIMO
\safemath{\altoutpvec}{\tilde{\outpvec}} % alternative output vector MIMO
\safemath{\wgnvec}{\vecw}				% noise MIMO
\safemath{\altwgnvec}{\tilde{\wgnvec}}
\safemath{\cminpvec}{\vecs}				% constant modulus input vector
\safemath{\cminp}{s}
\safemath{\cminpmat}{\matS}
\safemath{\tfoutpvec}{\bm{\mathsf{y}}}	% TF output vector
\safemath{\tfoutpvecp}{\tfoutpvec[\dtdf]}	% 	'' parametrized
\safemath{\alttfoutpvec}{\tilde{\tfoutpvec}}	% alternate TF output vector
\safemath{\alttfoutpvecp}{\alttfoutpvec[\dtdf]}	% 	'' parametrized
\safemath{\tfwgnvec}{\bm{\mathsf{w}}}		% TF noise vector
\safemath{\tfwgnvecp}{\tfwgnvec[\dtdf]}	% 	'' parametrized
\safemath{\alttfwgnvec}{\tilde{\tfwgnvec}}% alternate TF noise vector
\safemath{\alttfwgnvecp}{\alttfwgnvec[\dtdf]}	% 	'' parametrized
\safemath{\tfinpvec}{\bm{\mathsf{x}}}		% TF input vector
\safemath{\tfinpvecp}{\tfinpvec[\dtdf]}	% 	'' parametrized
\safemath{\alttfinpvec}{\tilde{\tfinpvec}}	% alternateTF input vector
\safemath{\alttfinpvecp}{\alttfinpvec[\dtdf]}	% 	'' parametrized
\safemath{\cmtfinpvec}{\bm{\mathsf{s}}}	% TF constant modulus input vector
\safemath{\coutcov}{\covmat{\outpvec\given\inpmat}}

\safemath{\Pave}{P}						% average power
\safemath{\avPnorm}{\Ex{}{\vecnorm{\inpvec}^{2}}/\tstep\leq\tslots\Pave}
\safemath{\peakPnorm}{\vecnorm{\tfinpvecp}^{2}/\tstep\le\papr\Pave/\fslots}
\safemath{\papr}{\beta}					% PAPR constraint

\safemath{\distset}{\setP}				% set of input distributions
\safemath{\distsetrel}{\setQ}			% extended set of input distributions
\safemath{\distsetres}{\distsetrel\rvert_{\avpopt}} % restr. input distr.

\safemath{\capacityp}{\capacity(\bandwidth)}
\safemath{\UBone}{U_1}				% first upper bound
\safemath{\UBonep}{\UBone(\bandwidth)}
\safemath{\UBc}{U_c}					% coherent upper bound
\safemath{\UBcp}{\UBc(\bandwidth)}
\safemath{\LBone}{L_1}				% first lower bound
\safemath{\LBonep}{\LBone(\bandwidth,\altnumtx)}
\safemath{\LBapprox}{L_a}				% first lower bound
\safemath{\LBapproxp}{\LBapprox(\bandwidth,\altnumtx)}
\safemath{\pt}{G_{\rxindex}}			% penalty term
\safemath{\ptp}{\pt(\bandwidth)}		% '' parametrized
\safemath{\avpopt}{\alpha}			% maximizing average power
\safemath{\avpoptp}{\alpha(\bandwidth)}	% parameterized ''
\safemath{\taylorone}{a}				% first-order Taylor coeff.
\safemath{\tsparam}{\gamma}			% time sharing parameter
\safemath{\rxevsqsum}{\theta}		% sum of squared RX correlation eigenvalues

\newcommand{\cex}[1]{e^{\iu2\pi #1}}		% positive complex exponential
\newcommand{\cexn}[1]{e^{-\iu2\pi #1}}	% negative complex exponential
\safemath{\limintime}{\lim_{\tslots\rightarrow\infty}} % limit in time
\safemath{\wpone}{\text{w.p.1}}			% with probability 1
\newcommand{\osmall}[1]{\landauo\lefto(#1\right)} % asymptotic notation
\safemath{\di}{N}						% generic dimension
\safemath{\indvar}{n}					% generic index
\safemath{\idx}{i}						% index variable
\newcommand{\dg}[1]{\mathrm{diag}\!\left\{#1\right\}}
\safemath{\evvec}{\veclambda}			% vector of eigenvalues
\newcommand{\spreadint}[1]{\iint_{\delay\,\doppler}#1d\doppler d\delay}
\newcommand{\covmat}[1]{\matR_{#1}}		% general covariance matrix
\safemath{\specparam}{\theta}
\safemath{\altspecparam}{\varphi}
\safemath{\chspec}{c}					% channel spectrum
\safemath{\chspecp}{\chspec(\specparam,\altspecparam)}
\safemath{\supPeakAveMT}{\sup_{\distset }}
\safemath{\imat}{\matI}					% identity matrix
\safemath{\zmat}{\bZero}				 % all-zero matrix
\safemath{\oppinvmat}{\tilde{\matA}}	% inverse diagonal matrix in App. A
\safemath{\setsize}{L}					% cardinality of the index set

\safemath{\fto}{g}
\safemath{\ftop}{\fto(\avpopt)}
\safemath{\ftoder}{\fto'}
\safemath{\ftoderp}{\ftoder(\avpopt)}

\newcommand{\pulsersii}{{{\sc Pulsers} Phase~II}\xspace}

\begin{document}
\IEEEoverridecommandlockouts
% DRAFT
% to include revision information into the resulting PDF
%\svnInfo $Id: ltv-mimo.tex 2346 2008-01-29 09:17:58Z gdurisi $

% add a PDFinfo field to store metadata in the output PDF
\pdfinfo{
	/Title		(Capacity Bounds for Peak-Constrained Multiantenna Wideband
				Channels)
	/Author		(Ulrich G. Schuster, Giuseppe Durisi, Helmut Boelcskei, and H. Vincent Poor)
	/Subject		(SVN revision 2340)
	/Keywords	(Noncoherent capacity, MIMO systems, underspread channels, wideband channels)
}

% paper title
\title{Capacity Bounds for Peak-Constrained Multiantenna Wideband Channels}
%
%
% author names and IEEE memberships
% note positions of commas and nonbreaking spaces ( ~ ) LaTeX will not break
% a structure at a ~ so this keeps an author's name from being broken across
% two lines.
% use \thanks{} to gain access to the first footnote area
% a separate \thanks must be used for each paragraph as LaTeX2e's \thanks
% was not built to handle multiple paragraphs
\author{Ulrich~G.~Schuster,~\IEEEmembership{Student Member,~IEEE},
	Giuseppe~Durisi,~\IEEEmembership{Member,~IEEE},
	Helmut~B\"olcskei,~\IEEEmembership{Senior Member,~IEEE},
	and~H.~Vincent~Poor,~\IEEEmembership{Fellow,~IEEE}%
\thanks{This work was supported partly by the European Commission through the
	Integrated Project~\pulsersii under contract No.~FP6-027142 and partly by
	the U.~S. National Science Foundation under Grants ANI-03-38807 and
	CNS-06-25637. Part of this work originated while U.~G.~Schuster was a
	visiting researcher at Princeton University. A conference version of this paper has been submitted to the IEEE Int. Symposium on Information Theory (ISIT), Toronto, Canada, July 2007.}%
\thanks{U.~G.~Schuster, G.~Durisi, and H.~B\"olcskei are with the Communication
	Technology Laboratory, ETH Zurich, 8092 Zurich, Switzerland (e-mail:
	\{schuster, gdurisi, boelcskei\}@nari.ee.ethz.ch).}%
\thanks{H.~V.~Poor is with Princeton University, Princeton, NJ~08544, U.S.A.
	(e-mail: poor@princeton.edu).}}

% make the title area
\maketitle

%%%%%%%%%%%%%%%%
\begin{abstract}
We derive bounds on the noncoherent capacity of a very general class of multiple-input
multiple-output channels that allow for selectivity in time and frequency
as well as for spatial correlation. The bounds apply to peak-constrained inputs;
they are explicit in the channel's scattering function, are useful for a large
range of bandwidth, and allow to coarsely identify the capacity-optimal
combination of bandwidth and number of transmit antennas. Furthermore, we obtain
a closed-form expression for the first-order Taylor series expansion of capacity
in the limit of infinite bandwidth. From this expression, we conclude that in
the wideband regime:
\begin{inparaenum}[(i)]
\item it is optimal to use only one transmit antenna when the channel is
	spatially uncorrelated;
\item rank-one statistical beamforming is optimal if the channel is spatially
	correlated; and
\item spatial correlation, be it at the transmitter, the receiver, or both, is
	beneficial.
\end{inparaenum}
\end{abstract}


%%%%%%%%%%%%%%%%%%%%%%%%%%%%%%%%%%%%%%%%%%%%%
\section{Introduction and Summary of Results}
\label{sec:introduction}

Bandwidth and space are sources of degrees of freedom that can be utilized to
transmit information over wireless fading channels. Channel measurements
indicate that an increase in the number of degrees of freedom also increases the
channel uncertainty that the receiver has to resolve~\cite{schuster07-07a}. If
the transmit signal is allowed to be peaky, that is, if it can have an unbounded
peak value, channel uncertainty is immaterial
in the limit of infinite bandwidth. Indeed, for a fairly general class of fading
channels, the capacity of the infinite-bandwidth additive white Gaussian noise
(AWGN)~channel can be achieved~\cite{gallager68a,telatar00-07a,durisi06-07a}.

A more realistic modeling assumption is to limit the peak power of the
transmitted signal. In this case, the capacity behavior of most channels changes
drastically: for certain types of peak constraints, the capacity can even
approach zero in the wideband
limit~\cite{telatar00-07a,medard02-04a,subramanian02-04a}. Intuitively, under a
peak constraint on the transmit signal, the receiver is no longer able to resolve the channel
uncertainty as the number of degrees of freedom increases. Consequently,
questions of significant practical relevance are how much bandwidth to use and
whether spatial degrees of freedom obtained by multiple antennas can be
exploited to increase capacity.

The aim of this paper is to characterize the capacity of spatially correlated multiple-input
multiple-output~(MIMO) fading channels that are time and frequency selective, 
i.e., that exhibit memory in frequency and time, given that
\begin{inparaenum}[(i)]
\item the transmit signal has bounded~peak power and
\item the transmitter and the receiver know the channel law but both are
	ignorant of the channel realization.\label{item:noncoherent-setting}
\end{inparaenum}
The assumptions~\eqref{item:noncoherent-setting} constitute the {\em noncoherent
setting}, as opposed to the {\em coherent setting} where the receiver has
perfect channel state information~(CSI) and the transmitter knows the channel
law only.

%--
\subsubsection*{Related Work}
Sethuraman~{\it et al.}~\cite{sethuraman08a} analyzed the capacity of
peak-constrained MIMO Rayleigh-fading channels that are frequency flat, time
selective, and spatially uncorrelated and derived an upper bound and a low-SNR
lower bound that allow to characterize the second-order Taylor series expansion
of capacity around the point $\text{SNR}=0$. In particular, it is shown in~\cite{sethuraman08a}
that in the low-SNR regime it is
optimal to use only a single transmit antenna, while additional receive antennas
are always beneficial. The low-SNR results also apply to a wideband channel
with fixed total transmit power and increasing bandwidth if the wideband channel
can be decomposed into a set of independent and identically distributed~(\iid)
parallel subchannels in frequency~\cite{sethuraman08a}.

Spatial correlation is often beneficial in the noncoherent setting. For the
{\em separable} (Kronecker) spatial correlation
model~\cite{chuah02-03a,kermoal02-08a}, Jafar and Goldsmith~\cite{jafar05-05a}
proved that transmit correlation increases the capacity of a memoryless fading
channel. Moreover, in the low-SNR regime, the rates achievable with on-off keying on
memoryless fading channels~\cite{zhang07-03a} and with finite-cardinality
constellations on block-fading channels~\cite{srinivasan07-10a} increase in the
presence of spatial correlation at the transmitter, the receiver, or both.

\subsubsection*{Contributions}

We consider a point-to-point MIMO~channel model where each {\em component
channel} between a given transmit antenna and a given receive antenna is {\em
underspread}~\cite{kennedy69} and satisfies the standard {\em wide-sense stationary
uncorrelated-scattering}~(WSSUS) assumption~\cite{bello63-12a}; hence, our channel model
allows for selectivity in time and frequency. We assume that the component
channels are spatially correlated according to the separable correlation
model~\cite{chuah02-03a,kermoal02-08a} and that they are characterized by the
same scattering function; furthermore, the transmit signal is peak constrained.
On the basis of a discrete-time, discrete-frequency approximation of said
channel model that is enabled by the underspread property~\cite{kozek97a}, we
obtain the following results:
\begin{itemize}
\item We derive upper and lower bounds on capacity. These bounds are explicit in
	the channel's scattering function and allow to coarsely identify the
	capacity-optimal combination of bandwidth and number of transmit antennas
	for a fixed number of receive antennas.
\item For spatially uncorrelated channels, we generalize the asymptotic results
	of Sethuraman~{\it et al.}~\cite{sethuraman08a} to time- {\em and}
	frequency-selective channels: for
	large enough bandwidth---or equivalently, for small enough SNR---it is
	optimal to use a single transmit antenna only, while additional receive
	antennas always increase capacity.
\item Differently from the coherent setting~\cite{jorswieck06-05a,tulino05-07a,lozano06a}, we find that
	both transmit {\em and} receive correlation are beneficial in the wideband
	regime. Furthermore, rank-one statistical beamforming along the
	strongest eigenmode of the spatial transmit correlation matrix is optimal
	for large bandwidth.
\end{itemize}

As the derivations of the results in the present paper rely on several
techniques developed in~\cite{durisi08a} for single-input single-output~(SISO)
time- and frequency-selective channels, we detail only the new elements in our
derivations and refer to~\cite{durisi08a} otherwise.

%--
\subsubsection*{Notation}

Uppercase boldface letters denote matrices and lowercase boldface letters
designate vectors. The superscripts~$\tp{}$, $\conj{}$, and~$\herm{}$ stand for
transposition, element-wise conjugation, and Hermitian transposition,
respectively. For two matrices~\matA and~\matB of appropriate dimensions, the
Hadamard product is denoted as~$\matA\had\matB$ and the Kronecker product is
denoted as~$\matA \kron\matB$; to simplify notation, we use the convention that
the ordinary matrix product always precedes the Kronecker and Hadamard products,
e.g., $\matA\matB\had\matC$ means~$(\matA\matB)\had\matC$ for some matrix~\matC
of appropriate dimension. We designate the identity matrix and the all-zero
matrix of dimension~$\fslots\times \fslots$ by~$\imat_{\fslots}$
and~$\zmat_{\fslots}$, respectively; $\sqm{\matD}$~is the unique nonnegative
definite square-root matrix of the nonnegative definite matrix~\matD. The
determinant of a square matrix~\matX is~$\det(\matX)$, its rank
is~$\rank(\matX)$, and its trace is~$\tr(\matX)$. The vector of eigenvalues
of~\matX is denoted by~$\evvec(\matX)$,  We let~$\dg{\vecx}$ denote a
diagonal square matrix whose main diagonal contains the elements of the
vector~\vecx. The function~$\dirac(x)$ is the Dirac distribution. All logarithms
are to the base~$e$. For two functions~$f(x)$ and~$g(x)$, the notation~$f(x) =
\osmall{g(x)}$ means that~$\lim_{x \to 0}f(x)/g(x)= 0$. If two random
variables~$a$ and~$b$ follow the same distribution, we write~$a\distas b$.
Finally, we denote the expectation operator by~$\Ex{}{\cdot}$ and the Fourier
transform operator by~$\four[\cdot]$.

%%%%%%%%%%%%%%%%%%%%%%
\section{System Model}
\label{sec:model}

In the following subsections, we first introduce the SISO~model for one
component channel and subsequently discuss the extension of this model to the
MIMO~setting.

%--------------------------------------
\subsection{Underspread WSSUS Channels}

The relation between the input signal~\inppct and the corresponding output 
signal~\outppct of a SISO stochastic linear time-varying~(LTV) channel~$\CHop$
can be expressed as
\ba
\label{eq:iorel-ct}
	\outppct=\bigl(\CHop\inp\bigr)(t)+ \wgn(\time)=\int_{\time'}
		\kernel(\time,\time')\inp(\time')d\time' + \wgn(\time)
\ea
where~$\kernel(\time,\time')$ denotes the random kernel of the channel
operator~$\CHop$ and~$\wgn(\time)$ is a white Gaussian noise process. We assume
that~\kernelp is a zero-mean jointly proper Gaussian~(JPG) process in~\time
and~$\time'$ whose Fourier transforms are well defined. In particular,
$\tvtf(\time,\freq) = \four_{\delay\to\freq}[\kernel(\time,\time-\delay)]$ is
called the {\em time-varying transfer function} and~$\spfp=\four_{\time\to
\doppler}[\kernel(\time,\time-\delay)]$ is called the {\em spreading function}.
We assume that the channel is~WSSUS, so that
\bas
	\Ex{}{\spfp\conj{\spf}(\doppler',\delay')}=\scfp\dirac(\doppler-\doppler')
		\dirac(\delay-\delay').
\eas
Consequently, the statistical properties of the channel~$\CHop$ are completely
specified through its so-called {\em scattering function}~\scfp. A~WSSUS channel
is said to be {\em underspread}~\cite{kozek97a} if~\scfp is compactly supported
on a rectangle~$[-\maxDoppler,\maxDoppler]\times[-\maxDelay,\maxDelay]$ whose
{\em spread}~$\spread\define4\maxDoppler\maxDelay$ satisfies~$\spread<1$.

%----------------------------------
\subsection{Discrete Approximation}
\label{sec:discrete-approx}

To simplify information-theoretic analysis, we would like to {\em diagonalize}
the channel operator~$\CHop$, i.e., replace the integral input-output~(IO) 
relation~\fref{eq:iorel-ct} by a {\em countable} set of {\em scalar}
IO~relations. To this end, we cannot use an eigendecomposition of
the random kernel~\kernelp because its eigenfunctions are random
as well, and hence unknown to the transmitter and the receiver in the
noncoherent setting. Yet, for underspread channels it is possible to find an
orthonormal set of {\em deterministic} approximate eigenfunctions that depend
only on the channel's scattering function~\cite{kozek97a}. Consequently,
knowledge of the channel law---and hence of the scattering function---is 
sufficient for transmitter and receiver
to approximately diagonalize~$\CHop$. One possible choice of approximate
eigenfunctions is the {\em Weyl-Heisenberg set} of mutually orthogonal
time-frequency shifts~$\logonp\define \logon(\time-\dtime\tstep)\cex{\dfreq
\fstep\time}$ of some prototype function~$\logon(\time)$ that is well localized
in time and frequency. The grid parameters~\tstep and~\fstep need to
satisfy~$\tfstep\ge1$; then, the kernel of~$\CHop$ can be approximated
as~\cite{durisi08a}
\ba 
\label{eq:approx-kernel}
	\kernelp \approx\sum_{\dtime=-\infty}^{\infty}\sum_{\dfreq=-\infty}^{\infty}
		\underbrace{\tvtf(\dtime\tstep,\dfreq\fstep)}_{\irp}\logonp\conj{\logon
		}_{\dtdf}(\time').
\ea
The approximation quality depends on the prototype function~$\logon(\time)$ and
on the parameters~\tstep and~\fstep, which need to be suitably chosen with
respect to the scattering function~\scfp~\cite{kozek97a,durisi08a}. The
eigenvalues of the approximate channel with kernel~\fref{eq:approx-kernel} are
given by~$\irp\define\tvtf(\dtime\tstep,\dfreq\fstep)$. As the channel is~JPG
and~WSSUS, the discretized channel process $\{\irp\}$ is also~JPG and stationary
in both discrete time~\dtime and discrete frequency~\dfreq. We denote its
correlation function by~$\chcorrp\define \Ex{}{\ir[\dtime'+\dtime,\dfreq'+
\dfreq]\conj{\ir}[\dtime',\dfreq']}$, normalized as~$\chcorr[0,0]=1$. The
associated spectral density
\bas
	\chspecp\define\sum_{\dtime=-\infty}^{\infty}
		\sum_{\dfreq=-\infty}^{\infty}\chcorr[\dtime,\dfreq]\cexn{(\dtime
		\specparam-\dfreq\altspecparam)},\quad\abs{\specparam},
		\abs{\altspecparam}\le 1/2
\eas
can be expressed in terms of the scattering function~\scfp as~\cite{durisi08a}
\ba\label{eq: spectral density}
	\chspecp = \frac{1}{\tfstep}\sum_{\dtime=-\infty}^{\infty}\sum_{\dfreq=
		-\infty}^{\infty}\scf\lefto(\frac{\specparam-\dtime}{\tstep},
		\frac{\altspecparam-\dfreq}{\fstep}\right).
\ea
We choose~$\tstep\le1/(2\maxDoppler)$ and~$\fstep\le1/(2\maxDelay)$ so that no
aliasing of the scattering function occurs in~\fref{eq: spectral density};
for this choice of~\tstep and~\fstep, the normalization~$\chcorr[0,0]=1$
implies that~$\int_{\delay}\!\int_{\doppler}{\scfp}d\doppler d\delay=1$.Ê
Next, we substitute the approximation~\fref{eq:approx-kernel}
into~\fref{eq:iorel-ct} and project the input signal~\inppct and the output
signal~\outppct onto the Weyl-Heisenberg set~$\{\logonp\}$ to obtain the
countable set of scalar IO~relations
\ba
\label{eq:iorel-dt-scalar}
	\outpp=\irp\inpp+\wgnp,
\ea
one for each time-frequency {\em slot}~$(\dtdf)$. The coefficients~$\{\wgnp\}$ are \iid JPG with zero mean and variance normalized to one.

%---------------------------------------------------------------
\subsection{Extension to Multiple Transmit and Receive Antennas}
\label{sec:mimo-extension}

We extend the SISO~channel model in~\eqref{eq:iorel-dt-scalar} to a MIMO~channel
model with~\numtx transmit antennas, indexed by~\txindex, and~\numrx receive antennas,
indexed by~\rxindex, and assume that all component channels are characterized by
the same scattering function~\scfp so that they are diagonalized by the same
Weyl-Heisenberg set~$\{\logonp\}$. For each slot~$(\dtdf)$ and component
channel~$(\rxindex,\txindex)$ the resulting scalar channel coefficient is
denoted as~$\ir_{\mimo}[\dtdf]$. We arrange the coefficients for a given
slot~$(\dtdf)$ in an~$\numrx\times\numtx$ matrix~$\tfchmatp$ with
entries~$[\tfchmatp]_{\rxindex,\txindex}=\ir_{\mimo}[\dtdf]$. The diagonalized
IO~relation of the multiantenna channel is then given by a countable set of
standard MIMO IO~relations of the form
\ba\label{eq:per-slot-io}
	\tfoutpvecp = \tfchmatp\tfinpvecp+\tfwgnvecp
\ea
where  $\tfinpvecp = \tp{\btm \inp_{0}[\dtdf] &
\inp_{1}[\dtdf] & \cdots & \inp_{\numtx-1}[\dtdf] \etm}$ is the
\numtx-dimensional input vector for each slot~$(\dtdf)$, $\tfoutpvecp = \tp{\btm \outp_{0}[\dtdf] & \outp_{1}[\dtdf] & \cdots & \outp_{\numrx-1}[\dtdf] \etm}$ is the
\numrx-dimensional output vector, and~\tfwgnvecp is the \numrx-dimensional noise
vector.\footnote{To distinguish quantities that pertain to the MIMO IO~relation
for an individual slot~$(\dtdf)$ from the corresponding quantities of the joint
time-frequency-space IO~relation~\fref{eq:iorel-mimo} to be introduced in the
next subsection, we use a sans-serif font for the former quantities.} We allow
for spatial correlation according to the separable correlation
model~\cite{chuah02-03a,kermoal02-08a}, so that
\bas
	\Ex{}{\ir_{\mimo}[\dtime'+\dtime,\dfreq'+\dfreq]\conj{\ir}_{\rxindex',
		\txindex'}[\dtime',\dfreq']} 
		= \rxcorr[\rxindex,\rxindex']\txcorr[\txindex,\txindex']\chcorrp.
\eas
The~$\numtx\times\numtx$ matrix~\txcorrmat with entries~$[\txcorrmat]_{\txindex,
\txindex'}=\txcorr[\txindex,\txindex']$ is called the {\em transmit correlation
matrix}, and the	$\numrx\times\numrx$ matrix~\rxcorrmat, with
entries~$[\rxcorrmat]_{\rxindex,\rxindex'}=\rxcorr[\rxindex,\rxindex']$, is the
{\em receive correlation matrix}. Consequently,
\ba
\label{eq:per-slot-correlated-channel}
	\tfchmatp = \sqm{\rxcorrmat}\tfchmatwp\tp{(\sqm{\txcorrmat})}
\ea
where~\tfchmatwp is an~$\numrx\times\numtx$
matrix with \iid JPG~entries of zero mean and unit variance for
all~$(\dtdf)$. We normalize~\txcorrmat and~\rxcorrmat so that~$\tr(\txcorrmat)=
\numtx$ and~$\tr(\rxcorrmat)=\numrx$.

%------------------------------------------------------------------------------
\subsection{Matrix-Vector Formulation of the Discretized Input-Output Relation}
\label{sec:mat-vec-notation}

We define a {\em channel use} as a~$\tslots\times\fslots$ rectangle of
time-frequency slots and stack the symbols~$\{\inp_{\txindex}[\dtdf]\}$
transmitted from all~\numtx transmit antennas during one channel use into an
\txdim-dimensional vector~\inpvec, the corresponding output~$\{\outp_{\rxindex}
[\dtdf]\}$ for all~\numrx receive antennas into an \rxdim-dimensional
vector~\outpvec, and likewise the noise~$\{\wgn_{\rxindex}[\dtdf]\}$ into an
\rxdim-dimensional vector~\wgnvec. Stacking proceeds first along frequency, then
along time, and finally along space, as shown exemplarily for the input
vector~\inpvec:
\begin{subequations}
\label{eq:stacking}
\ba
	\inpvec_{\txindex}[\dtime] &= \tp{[\inp_{\txindex}[\dtime,0]\,
		\inp_{\txindex}[ \dtime,1]\, \cdots\, 
		\inp_{\txindex}[\dtime,\fslots-1]]}\\
	\inpvec_{\txindex}&=\tp{[\tp{\inpvec}_{\txindex}[0]\,
		\tp{\inpvec}_{\txindex}[1]\, \cdots\,
		\tp{\inpvec}_{\txindex}[\tslots-1]]}\label{eq:inpvec-per-antenna}\\
	\inpvec &= \tp{[\tp{\inpvec}_{0}\, \tp{\inpvec}_{1}\,
		\cdots\, \tp{\inpvec}_{\numtx-1}]}.
\ea
\end{subequations}
Analogously, we stack the channel coefficients, first in frequency to obtain
the vectors~$\chvec_{\mimo}[\dtime]$, and then in time to obtain a
vector~$\chvec_{\mimo}$ for each component channel~$(\rxindex,\txindex)$;
further stacking of these vectors along transmit antennas~\txindex and then
along receive antennas~\rxindex results in the~\totdim-dimensional
vector~\chvec. Let~$\inpmat_{\txindex}=\dg{\inpvec_{\txindex}}$ and $\inpmat =
[\inpmat_{0}\, \inpmat_{1}\, \cdots\, \inpmat_{\numtx-1}]$, where the
vectors~$\inpvec_{\txindex}$ are defined in~\fref{eq:inpvec-per-antenna}.
With this notation, the IO~relation for one channel use can be
conveniently expressed as
\ba
\label{eq:iorel-mimo}
	\outpvec=  (\imat_{\numrx} \kron \inpmat)\chvec + \wgnvec.
\ea
The distribution of the channel coefficients in a given channel use is
completely characterized by the~$\totdim\times\totdim$ correlation matrix
\ba
\label{eq:chcorr}
	\Ex{}{\chvec\herm{\chvec}} = \rxcorrmat\kron\txcorrmat\kron\schcorrmat
\ea
where the correlation matrix~$\schcorrmat\define\Exop[\chvec_{\mimo}
\herm{\chvec}_{\mimo}]$ is the same for all component channels~$(\rxindex,
\txindex)$ by assumption; \schcorrmat~is two-level Toeplitz, i.e.,
block-Toeplitz with Toeplitz blocks. We assume that the three
matrices~\txcorrmat, \rxcorrmat, and~\schcorrmat are known to the transmitter
and the receiver.

%-----------------------------
\subsection{Power Constraints}
\label{sec:power-constraints}

We impose a constraint on the average power of the transmitted signal per
channel use such that~\avPnorm. In addition, we assume a peak constraint
across transmit antennas in each slot~(\dtdf) according to:
\ba
\label{eq:peak-sum}
	\frac{1}{\tstep}\sum_{\txindex=0}^{\numtx-1}\abs{\inp_{\txindex}[\dtdf]}^{2}\leq \frac{\papr\Pave}{\fslots}
\ea
with probability~1 (\wpone). Here,~$\papr\geq 1$ is the peak- to average-power ratio (PAPR).

%--------------------------------------------------------
\subsection{Spatially Decorrelated Input-Output Relation}
\label{sec:spatially-decorrelated}

Before proceeding to analyze the capacity of the channel just introduced,
we make one more cosmetic change to the IO~relation~\fref{eq:iorel-mimo}, 
which simplifies the exposition of our results considerably. For each slot, we
express the input and output vectors in the coordinate systems defined by the
eigendecomposition of the transmit and receive correlation matrices,
respectively. A~similar transformation is used
in~\cite{jafar05-05a, srinivasan07-10a} for a frequency-flat block-fading
spatially correlated MIMO~channel. Let the eigendecomposition of the spatial
correlation matrices be
\begin{subequations}
\bas
	\txcorrmat = \txcorrunit\txcorrspec\herm{\txcorrunit}, \qquad\qquad
	\rxcorrmat &= \rxcorrunit\rxcorrspec\herm{\rxcorrunit},
\eas
\end{subequations}
where~$\txcorrspec=\dg{\tp{[\txev_{0}\, \txev_{1}\,\cdots\,\txev_{\numtx-1}]}}$
contains the eigenvalues~$\{\txev_{\txindex}\}$ of~\txcorrmat, ordered according
to $\txev_{0}\geq\txev_{1}\geq\cdots\geq\txev_{\numtx-1}$ and, similarly,
$\rxcorrspec=\dg{\tp{[\rxev_{0}\, \rxev_{1}\,\cdots\,\rxev_{\numrx-1} ]}}$
contains the eigenvalues~$\{\rxevp\}$ of~\rxcorrmat, ordered according to
$\rxev_{0}\geq\rxev_{1}\geq\cdots\geq\rxev_{\numrx-1}$. The columns
of~\txcorrunit are called the {\em transmit eigenmodes} and the columns
of~\rxcorrunit are the {\em receive eigenmodes}. Instead of the
vectors~\tfinpvecp and~\tfoutpvecp, we use the rotated vectors~$\tp{\txcorrunit}\tfinpvecp$ and~$\herm{\rxcorrunit}\tfoutpvecp$, respectively, to obtain the following {\em spatially decorrelated} IO~relation in each slot~$(\dtdf)$:
\be
\bs
	\herm{\rxcorrunit}&\tfoutpvecp = \herm{\rxcorrunit}\tfchmatp\tfinpvecp
		+ \herm{\rxcorrunit}\tfwgnvecp\\
	&\stackrel{(a)}{=}\herm{\rxcorrunit}\bigl(\rxcorrunit\sqm{\rxcorrspec}
		\herm{\rxcorrunit}\bigr)\tfchmatwp\tp{\bigl(\txcorrunit\sqm{\txcorrspec}
		\herm{\txcorrunit}\bigr)}\tfinpvecp\\
		&\hphantom{\stackrel{(a)}{=}\herm{\rxcorrunit}\bigl(\rxcorrunit
			\sqm{\rxcorrspec}\herm{\rxcorrunit}\bigr)\tfchmatwp\txcorrunit
			\sqm{\txcorrspec}\herm{\txcorrunit}}
			+\herm{\rxcorrunit}\tfwgnvecp\\
	&=\sqm{\rxcorrspec}\herm{\rxcorrunit}\tfchmatwp\conj{\txcorrunit}
		\sqm{\txcorrspec}\tp{\txcorrunit}\tfinpvecp + \herm{\rxcorrunit}
		\tfwgnvecp
\es
\label{eq:spatial-decorrelation}
\ee
where~(a) follows from~\fref{eq:per-slot-correlated-channel}. Rotations are
unitary operations; therefore,~$\herm{\rxcorrunit}\tfchmatwp\conj{\txcorrunit}\distas\tfchmatwp$ and~$\herm{\rxcorrunit}\tfwgnvecp\distas\tfwgnvecp$.
Furthermore, rotations preserve norms, so that the rotated input
vector~$\tp{\txcorrunit}\tfinpvecp$ satisfies the same power constraints as the
unrotated input vector~\tfinpvecp. Finally, $\herm{\rxcorrunit}\tfoutpvecp$ is a
sufficient statistic for the output vector~$\tfoutpvecp$. These three properties
imply that the capacity of the channel with input~\tfinpvecp and
output~\tfoutpvecp in~\eqref{eq:per-slot-io} is the same as the capacity of the
spatially decorrelated channel~$\sqm{\rxcorrspec}	\tfchmatwp\sqm{\txcorrspec}$
in~\eqref{eq:spatial-decorrelation} with input~$\tp{\txcorrunit}\tfinpvecp$ and
output~$\herm{\rxcorrunit}\tfoutpvecp	$. In the new coordinate system, \txindex
indexes transmit eigenmodes instead of transmit antennas, and~\rxindex indexes
receive eigenmodes instead of receive antennas.

It is now tedious but straightforward to similarly rotate the stacked
IO~relation~\fref{eq:iorel-mimo}. To keep notation simple, we chose not to
introduce new symbols for the rotated input and output and for the spatially
decorrelated channel; from here on, all inputs and
outputs are with respect to the rotated coordinate systems,
and the channel vector~\chvec now stands for the spatially decorrelated stacked
channel with correlation matrix
\ba
\label{eq:chcorr-sd}
	\Ex{}{\chvec\herm{\chvec}} = \rxcorrspec\kron\txcorrspec\kron\schcorrmat.
\ea
This correlation matrix is block diagonal, and hence of much simpler structure
than~\fref{eq:chcorr}.

%---------------------------------------------------
\subsection{Advantages and Limitations of the Model}

The channel model just presented is fairly general: it allows for correlation in
space and for selectivity in time and frequency. Hence, we can dispense with the
often used block-fading assumption in time and with the assumption of
independent subchannels in frequency. Fortunately, the generality of our model
does not come at the price of high modeling complexity as only the scattering
function and the spatial correlation matrices~\txcorrmat and~\rxcorrmat are
needed to describe the distribution of the channel coefficients~$\{\ir_{\mimo}[\dtdf]\}$.
Both the scattering function and the spatial correlation matrices can be
obtained from channel measurements~\cite{cox72-09a,artes04-05a,kermoal02-08a},
so that the model can be directly related to real-world channels.

Modeling is synonymous with making assumptions and simplifications. We briefly 
discuss and justify our key assumptions.
\begin{compactitem}
\item The assumption that transmitter and receiver do not know the channel
	realization is accurate, as in a practical system channel realizations
	can only be inferred from the received signal. The rates achievable with
	specific methods to obtain~CSI, like training schemes, cannot exceed the
	capacity of the channel in the noncoherent setting.
\item Virtually all wireless channels are highly underspread: extremely
	dispersive outdoor channels with fast moving terminals may have a 
	spread of~$\spread\approx10^{-2}$, while for slowly varying indoor channels
	typically~$\spread\approx10^{-7}$.
\item The Weyl-Heisenberg transmission set~$\{\logonp\}$ can be interpreted as
	pulse-shaped~(PS) orthogonal frequency-division multiplexing~(OFDM); hence,
	the model we use in our information-theoretic analysis is directly related
	to a practical transmission scheme.
\item We neglect the error incurred by the approximation of the kernel~\kernelp
	in~\fref{eq:approx-kernel}, which is equivalent to neglecting intersymbol
	and intercarrier interference in the corresponding PS-OFDM system
	interpretation~\cite{durisi08a}. Yet, if the pulse~$\logon(\time)$
	and~\tstep and~\fstep are chosen so as to optimally mitigate intersymbol and
	intercarrier interference, i.e., if they are {\em matched} to the channel's
	scattering function~\cite{kozek97a,matz07-05a,durisi08a}, we conjecture that
	the resulting approximation error in~\fref{eq:approx-kernel} is smaller than
	the corresponding error incurred if either conventional cyclic prefix~OFDM
	or direct sampling of~\kernelp and truncation of the resulting sample
	sequence (e.g., see~\cite{medard02-04a}) is used to analyze underspread
	WSSUS~channels. In fact, these last two decompositions are, in general, not matched to the 
	channel's scattering function. 
\item The scattering function models small-scale fading, i.e., the statistical
	variation of the channel as transmitter, receiver, or objects in the
	propagation environment are displaced by a few wavelengths~\cite{tse05a}.
	Therefore, if the antennas at each terminal are spaced only a few
	wavelengths apart, the component channels may be well modeled by the same
	scattering function.
\item  We assume that the component channels are spatially correlated according
	to the separable correlation model~\cite{chuah02-03a,kermoal02-08a}. This
	assumption is common in theoretical analyses of MIMO~channels because it
	greatly simplifies analytical developments. Shortcomings of this model are
	discussed in~\cite{ozcelik03-08a,weichselberger06-01a}.
\item We assume that spatial correlation does not change over time and
	frequency. This assumption is valid only over a limited time duration and
	bandwidth, as it requires the antenna patterns to be constant over frequency
	and the configuration of dominant scattering clusters to be constant over
	time.
\item The constraint on the peak power across antennas is a reasonable
	model for a regulatory limit on the total isotropic radiated peak power. If
	the peak limitation arises from the power amplifiers in the individual
	transmit chains, a peak constraint per antenna should be used instead.
\end{compactitem}

%%%%%%%%%%%%%%%%%%%%%%%%%
\section{Capacity Bounds}
\label{sec:bounds}

With the system model and power constraints in place, we can now proceed to
evaluate upper and lower bounds on the capacity of the channel with
IO~relation~\fref{eq:iorel-mimo}. Although all results to follow pertain to the
channel model described in \fref{sec:mat-vec-notation} under the power
constraints in \fref{sec:power-constraints}, we use the spatially decorrelated
channel and the rotated input and output vectors introduced in
\fref{sec:spatially-decorrelated} to simplify the exposition of the proofs.

As we assume that for all~$(\rxindex,\txindex)$ the process~$\{\ir_{\mimo}[\dtdf]\}$ has a spectral density, given in~\eqref{eq: spectral density},
$\{\ir_{\mimo}[\dtdf]\}$ is ergodic in~\dtime for all component
channels~\cite{maruyama49a}, and the capacity is given
by~\cite[Chapter 12]{gray07b}
\ba
\label{eq:channel-capacity}
	\capacityp = \limintime\frac{1}{\tslots\tstep}\supPeakAveMT
		\mi(\outpvec;\inpvec)
\ea
for any fixed bandwidth~$\bandwidth=\fslots\fstep$. The supremum is taken over
the set~\distset of all input distributions that satisfy the constraints on peak
and average power in \fref{sec:power-constraints}.

%-----------------------
\subsection{Upper Bound}
\label{sec:ub}

%% Theorem upper bound
\begin{thm}
\label{thm:ubTFpeak}
The capacity~\fref{eq:channel-capacity} of the underspread WSSUS MIMO channel
in \fref{sec:mat-vec-notation} under the power constraints in \fref{sec:power-constraints} is upper-bounded as~$\capacity(\bandwidth)\leq\UBonep$, where
\begin{subequations}
\label{eq:ubone}
\ba
	\UBonep&=\notag\\
	\sup_{0\le\avpopt\le\txevm}&\sumz{\rxindex}{\numrx}
			\Biggl(\frac{\bandwidth}{\tfstep}\log\Bigl(1+\avpopt
		\rxevp\frac{\Pave\tfstep}{\bandwidth}\Bigr)
		-\avpopt\ptp\Biggr)\label{eq:ubTFpeak-core}\\
	\ptp&=\frac{\bandwidth}{\txevm\papr}\spreadint{\log\lefto(1+\frac{
		\txevm\rxevp\papr\Pave}{\bandwidth}\scfp\right)}.
		\label{eq:ubTFpeak-pt}
\ea
\end{subequations}
\end{thm}
\begin{IEEEproof}
Let~\distsetrel be the set of input distributions that satisfy
\ba
\label{eq:avp-relaxed}
	\frac{1}{\tstep}\Ex{}{\sumz{\txindex}{\numtx}\txev_{\txindex}\vecnorm{
		\inpvec_{\txindex}}^{2}}\le\txevm\tslots\Pave
\ea
and the peak constraint~\fref{eq:peak-sum}. As $\sumz{\txindex}{\numtx}
\txev_{\txindex}\Ex{}{\vecnorm{\inpvec_{\txindex}}^{2}}\le\txevm\sumz{\txindex}
{\numtx}\Ex{}{\vecnorm{\inpvec_{\txindex}}^{2}}=\txevm\Ex{}{\vecnorm{\inpvec}^{2}}$, any input distribution that
satisfies the average-power constraint~\avPnorm also
satisfies~\eqref{eq:avp-relaxed}, so that~$\distset\subset\distsetrel$. To
upper-bound~$\capacity(\bandwidth)$, we replace the supremum over~\distset
in~\fref{eq:channel-capacity} with a supremum over~\distsetrel and then use
the chain rule for mutual information and split the supremum over \distsetrel:
\bml
	\supPeakAveMT\mi(\outpvec;\inpvec)\le\sup_{\distsetrel}\mi(\outpvec;
		\inpvec)\\
	\le\sup_{0\le\avpopt\le\txevm}\Bigl\{\sup_{\distsetres}
		\mi(\outpvec;\inpvec,\chvec)-\inf_{\distsetres}\mi(\outpvec;\chvec\given
		\inpvec) \Bigr\}
\label{eq:decomposition}
\eml
where the distributions in the restricted set~\distsetres satisfy the equality
constraint~$\Exop\bigl[\sumz{\txindex}{\numtx}\txev_{\txindex}
\vecnorm{\inpvec_{\txindex}}^{2}\bigr]=\avpopt\tslots\Pave\tstep$
and the peak constraint~\fref{eq:peak-sum}.

To upper-bound~$\sup_{\distsetres}\mi(\outpvec;\inpvec,	\chvec)$, we drop the
peak constraint and take~$(\imat_{\numrx} \kron\inpmat)\chvec$ as~JPG distributed with block-diagonal correlation matrix~$\rxcorrspec\kron\Ex{}
{\inpmat(\txcorrspec\kron\schcorrmat)\herm{\inpmat}}$. Then,
\be
\bs
	\mi(\outpvec;&\inpvec,\chvec)\\
	&\stackrel{(a)}{\leq} \sumz{\rxindex}{\numrx}
		\log\det\Bigl(\imat_{\tfslots}
		+\rxevp\sumz{\txindex}{\numtx}\txev_{\txindex}\Ex{}
		{\inpvec_{\txindex}\herm{\inpvec}_{\txindex}}\had\schcorrmat\Bigr)\\
	&\stackrel{(b)}{\leq}\sumz{\rxindex}{\numrx}\sumz{\dfreq}{\fslots}
		\sumz{\dtime}{\tslots}\log\Bigl(1 + \rxevp
		\sumz{\txindex}{\numtx}\txev_{\txindex}\Exop\bigl[\abs{\inp_{\txindex}
		[\dtdf]}^{2}	\bigr]\Bigr)\\
	&\stackrel{(c)}{\le}\tfslots\sumz{\rxindex}{\numrx}\log\lefto(1 +
		\frac{\avpopt\rxevp\Pave\tstep}{\fslots}\right).
\es
\label{eq:ub-term1}
\ee
Here, (a)~follows from the assumption that~$(\imat_{\numrx} \kron\inpmat)\chvec$
is JPG distributed, from the block diagonal structure of its
correlation matrix, and because~$\inpmat(\txcorrspec\kron\schcorrmat)
\herm{\inpmat} = \sumz{\txindex}{\numtx} \txev_{\txindex} \inpmat_{\txindex}
\schcorrmat\herm{\inpmat_{\txindex}}=\sumz{\txindex}{\numtx}\txev_{\txindex}\inpvec_{\txindex}
\herm{\inpvec}_{\txindex}\had\schcorrmat$. Hadamard's inequality and the
normalization~$\chcorr[0,0]=1$ give~(b); finally, (c)~follows from Jensen's inequality.

The derivation of a lower bound on~$\inf_{\distsetres}\mi(\outpvec;\chvec\given
\inpvec)$ is more involved. Our proof is similar to the proof of the
corresponding SISO~result in~\cite[Theorem 1]{durisi08a}; therefore, we
highlight the novel steps only:
\be
\bs
	\inf_{\distsetres}&\mi(\outpvec;\chvec\given\inpvec)\\
	&\stackrel{(a)}{=}\inf_{\distsetres}\sumz{\rxindex}{\numrx}\Ex{}{
		\logdet{\imat_{\tfslots}+\rxevp\inpmat(\txcorrspec\kron
		\schcorrmat)\herm{\inpmat}}}\\
	&\stackrel{(b)}{=}\inf_{\distsetres}\sumz{\rxindex}{\numrx}\Exop
		\Biggl[\left(\frac{\logdet{\imat_{\tfslots}+\rxevp\inpmat(\txcorrspec
		\kron\schcorrmat)\herm{\inpmat}}}{\sumz{\txindex}{\numtx}
		\txev_{\txindex}	\vecnorm{\inpvec_{\txindex}}^{2}}\right)\\
		&\hphantom{\stackrel{(b)}{=}\inf_{\distsetres}\sumz{\rxindex}{\numrx}
		\Exop\Biggl[}
		\times\left(\sumz{\txindex}{\numtx}\txev_{\txindex}		\vecnorm{\inpvec_{\txindex}}^{2}\right)\Biggr]\\
	&\stackrel{(c)}{\ge} \sumz{\rxindex}{\numrx}\inf_{\inpvec}\frac{\log\det
		\bigl(\imat_{\tfslots}+\rxevp\sumz{\txindex}{\numtx}
		\txev_{\txindex}\inpvec_{\txindex}\herm{\inpvec}_{\txindex}
		\had\schcorrmat\bigr)}{\sumz{\txindex}{\numtx}\txev_{\txindex}
		\vecnorm{\inpvec_{\txindex}}^{2}}\\
		&\hphantom{\stackrel{(b)}{=}\inf_{\distsetres}\sumz{\rxindex}{\numrx}
		\Exop\Biggl[}
		\times\inf_{\distsetres}\Ex{}{\sumz{\txindex}{\numtx}
		\txev_{\txindex}	\vecnorm{\inpvec_{\txindex}}^{2}}\\
	&\stackrel{(d)}{\ge}\avpopt\tslots\Pave\tstep\sumz{\rxindex}{\numrx}
		\inf_{\inpvec}\frac{\log\det\bigl(\imat_{\tfslots}+\rxevp\!\!
		\sum\limits_{\txindex=0}^{\numtx-1}\txev_{\txindex}
		\herm{\inpmat}_{\txindex}\inpmat_{\txindex}\schcorrmat\bigr)}
		{\sumz{\txindex}{\numtx}\txev_{\txindex}
		\vecnorm{\inpvec_{\txindex}}^{2}}\\
	&\stackrel{(e)}{\ge}\frac{\avpopt\tslots\tstep\bandwidth}{\txevm\papr}
		\sumz{\rxindex}{\numrx}\spreadint{\log\Bigl(1+\frac{\txevm
		\rxevp\papr\Pave}{\bandwidth}\scfp\Bigr)}.
\es
\notag
\ee
Here, (a)~follows from the block-diagonal structure
of~$\rxcorrspec\kron\inpmat(\txcorrspec\kron\schcorrmat)\herm{\inpmat}$; to
obtain~(b), we multiply and divide by $\sumz{\txindex}{\numtx}\txev_{\txindex}
\vecnorm{\inpvec_{\txindex}}^{2}$, and to get (c) we replace the first factor in
the expectation by its infimum over all input vectors that satisfy the peak
constraint~\fref{eq:peak-sum}; (d)~follows because~$\Exop\bigl[\sumz{\txindex}
{\numtx}\txev_{\txindex}\vecnorm{\inpvec_{\txindex}}^{2}\bigr]=\avpopt\tslots
\Pave\tstep$ and because~$\det(\imat_{\di}+\matA\had\matB)\ge\det(\imat_{\di}+
(\imat_{\di}\had\matA)\matB)$ for two $\di\times\di$ nonnegative definite matrices~\matA and~\matB---a determinant inequality that we prove in
\fref{app:det-inequality}; finally, (e)~is a
consequence~\cite[Appendix B]{durisi08a} of the relation between mutual
information and minimum mean square estimation error~\cite{guo05-04a}. To
conclude the proof, we note that the bounds on both terms on the right-hand
side~(RHS) of~\eqref{eq:decomposition} no longer depend on~\tslots upon
division by~$\tslots\tstep$.
\end{IEEEproof}

%--
\subsubsection{The Supremum of~\UBonep}
\label{sec:ubopt}
As the value of~\avpopt that achieves the supremum in~\eqref{eq:ubTFpeak-core}
depends on~\bandwidth in general, the upper bound~\UBonep is difficult to
interpret. However, for the special case that the supremum is attained
for~$\avpopt=\txevm$ independently of~\bandwidth, the upper bound can be
interpreted as the capacity of a set of~\numrx parallel AWGN~channels with
received power~$\txevm\rxevp\Pave$ and~$\bandwidth/(\tfstep)$ degrees of freedom
per second, minus a penalty term that quantifies the capacity loss because of
channel uncertainty. We show in \fref{app:avpopt} that
a sufficient condition for the supremum in~\eqref{eq:ubTFpeak-core} 
to be achieved for~$\avpopt=\txevm$ is
\begin{subequations}
\label{eq:avpopt-conditions}
\ba
\label{eq:avpopt-spread}
	\spread&\le{\papr}/{(3\tfstep)}
\intertext{and}
\label{eq:avpopt-snr}
	0\le\frac{\Pave}{\bandwidth}&<\frac{\spread}{\txevm\rxevm\papr}
		\left[\exp\lefto(\frac{\papr}{2\tfstep\spread}\right)-1\right].
\ea
\end{subequations}
As virtually all wireless channels are highly underspread, as~$\papr\geq 1$
and, typically,~$\tfstep\approx 1.25$, condition~\eqref{eq:avpopt-spread}
is always satisfied, so that the only relevant condition is~\fref{eq:avpopt-snr};
but even for large channel spread, this condition holds for all
SNR~values~$\Pave/\bandwidth$ of practical interest. As an example, consider a
system with~$\papr=1$, and~$\numtx=\numrx=4$ that operates over a
channel with spread~$\spread=10^{-2}$. If we use the upper
bound~$\txevm\rxevm\le\numrx\numtx$, which follows from the
normalization~$\tr(\txcorrmat)=\numtx$ and~$\tr(\rxcorrmat)=\numrx$, we find
from~\fref{eq:avpopt-conditions} that~$\Pave/\bandwidth< 141\dB$ is sufficient 
for the supremum in~\eqref{eq:ubTFpeak-core} to be achieved for~$\avpopt=\txevm$.
This value is far in excess of the receive~SNR encountered in practical systems. Therefore,
we exclusively consider the case~$\avpopt=\txevm$ in the remainder of the
paper.

%--
\subsubsection{The Penalty Term}
What we call the ``penalty term'', i.e.,~$\txevm\sumz{\rxindex}{\numrx}\ptp$
in~\fref{eq:ubone}, is a lower bound on~$\inf_{\distsetres}\mi(\outpvec;\chvec
\given\inpvec)$. For SISO~channels, it is shown in~\cite{durisi08a} that of all
unit-volume scattering functions with prescribed~\maxDoppler and~\maxDelay, the brick-shaped scattering function, $\scfp=1/\spread$ for~$(\dd)\in[-\maxDoppler,
\maxDoppler]\times[-\maxDelay,\maxDelay]$, results in the largest penalty term. The
same is true for the MIMO~channel at hand, where the corresponding capacity 
is upper-bounded as
\bml
\label{eq: worst case upper bound}
	\capacityp\le  \sumz{\rxindex}{\numrx} 
	    \biggl\{\frac{\bandwidth}{\tfstep}\log\lefto(1+\txevm\rxevp
		\frac{\Pave\tfstep}{\bandwidth}\right) \\
		-\frac{\bandwidth\spread}{\papr}\log\lefto(1+\txevm\rxevp\frac{\papr
		\Pave}{\bandwidth\spread}\right)\biggr\}.
\eml
The upper bound~\eqref{eq: worst case upper bound} depends on the channel
spread~\spread and the PAPR~\papr only through their ratio, so that a
decrease in~\spread has the same effect on the upper bound as an increase in the
PAPR~\papr of the input signal.

\subsubsection{Spatial Correlation and Number of Antennas}

The upper bound~\UBonep depends on the transmit correlation matrix~\txcorrmat
only through its maximum eigenvalue~\txevm, which plays the role of a power
gain. This observation shows that rank-one statistical beamforming along any
eigenvector of~\txcorrmat corresponding to~\txevm is optimal whenever~\UBonep is
tight. At high~$\Pave/\bandwidth$ and correspondingly small bandwidth, \UBonep
increases linearly in the number of nonzero eigenvalues of the receive
correlation matrix, that is, in~$\rank(\rxcorrmat)$. As the capacity in the
coherent setting, which is a simple upper bound on~$\capacityp$, increases at
high~$\Pave/\bandwidth$ linearly only in the minimum of~$\rank(\txcorrmat)$
and~$\rank(\rxcorrmat)$~\cite[Proposition 4]{tulino05-07a}, we conclude
that~\UBonep is not tight at high~$\Pave/\bandwidth$. However, for large
bandwidth and corresponding small~$\Pave/\bandwidth$, we show in
\fref{sec:wideband-asymptotes} that~\UBonep is tight and that rank-one
statistical beamforming is indeed optimal in the wideband regime.

%----------------------
\subsection{Lower Bound}
\label{sec:lb}

\begin{thm}
\label{thm:lb}

Let~\chspecmatp denote the  $\fslots\times\fslots$ matrix-valued spectral
density of an arbitrary component channel\footnote{The vector
processes~$\chvec_{\mimo}[\dtime]$ of all component channels~$(\rxindex,
\txindex)$ have the same spectral density by assumption; therefore, we drop the
subscripts~\rxindex and~\txindex.} $\{\chvec[\dtime]\}$, i.e.,
\bas
	\chspecmatp \define \sum_{\dtime=-\infty}^{\infty}\Ex{}{\chvec[\dtime'+
		\dtime]\herm{\chvec}[\dtime']}\cexn{\dtime\specparam},\quad
		\abs{\specparam}	\le\frac{1}{2}.
\eas
Furthermore, let~\cmtfinpvec denote an~\numtx-dimensional vector whose
first~\altnumtx elements are \iid and of {\em constant modulus}---they have zero
mean and satisfy $\abs{[\cmtfinpvec]_{\txindex}}^{2}= \Pave \tstep/(\altnumtx\fslots
)$---and let the remaining~$\numtx-\altnumtx$ elements be zero.
Let~\tfchmatw be an $\numrx\times\numtx$ matrix and let~\tfwgnvec be an
\numrx-dimensional vector, both with \iid JPG~entries of zero mean and unit
variance.  Finally, denote by~$\mi(\tfoutpvec;\cmtfinpvec\given\tfchmatw)$ the
coherent mutual information of the memoryless fading MIMO~channel with
IO~relation~$\tfoutpvec=\sqm{\rxcorrspec}\tfchmatw\sqm{\txcorrspec}
\cmtfinpvec + \tfwgnvec$. Then, the capacity~\fref{eq:channel-capacity} of the
underspread WSSUS~MIMO channel in \fref{sec:mat-vec-notation} under the power constraints in \fref{sec:power-constraints} is lower-bounded as~$\capacity(
\bandwidth)\ge\max_{1\le \altnumtx \le \numtx}\LBonep$, where
\begin{multline}
\label{eq:lb}
	\LBonep=\max_{1\le\tsparam\le\papr}\Biggl\{\frac{\bandwidth}
		{\tsparam\tfstep}\mi(\tfoutpvec;\sqrt{\tsparam}\cmtfinpvec\given
		\tfchmatw) \\
	- \frac{1}{\tsparam\tstep}\sumz{\txindex}	{\altnumtx}\sumz{\rxindex}{\numrx}
		\!\int_{-1/2}^{1/2}\!\!\logdet{\imat_{\fslots}+\txev_{\txindex}
		\rxevp\frac{\tsparam\Pave\tfstep}{\altnumtx\bandwidth}
		\chspecmatp\!}\!	d\specparam \Biggr\}.
\end{multline}
\end{thm}

\begin{IEEEproof}
Any specific input distribution leads to a lower bound on capacity; in
particular, we choose to transmit constant modulus
symbols~$\inp_{\txindex}[\dtdf]=\cminp_{\txindex}[\dtdf]$ that are \iid over time, frequency, and
eigenmodes, and that satisfy~$\abs{\cminp_{\txindex}[\dtdf]}^{2}=\Pave\tstep/
(\altnumtx\fslots)$~\wpone for all~\dtdf and for~$\allz{\txindex}{\altnumtx}$.
The remaining~$\numtx-\altnumtx$ eigenmodes are not used to transmit
information. We stack the symbols $\cminp_{\txindex}[\dtdf]$ as
in~\fref{eq:stacking} and define the $\tfslots\times \txdim$ matrix
\bas
	\cminpmat \define \mat\cminpmat_{0} & \cminpmat_{1} & \cdots &
		\cminpmat_{\altnumtx-1} & \zmat_{\tfslots} & \cdots & \zmat_{\tfslots}
		\emat
\eas
with~$\cminpmat_{\txindex}\define\dg{\cminpvec_{\txindex}}$ and where the
last~$\numtx-\altnumtx$ entries are all-zero matrices~$\zmat_{\tfslots}$. Next, we use
\ba
\label{eq:lb-mi-decomposition}
	\mi(\outpvec;\cminpvec) \ge \mi(\outpvec;\cminpvec\given\chvec) -
		\mi(\outpvec;\chvec\given\cminpvec)
\ea
and bound the two terms on the RHS of~\eqref{eq:lb-mi-decomposition} separately. Because the input is~\iid,
$\mi(\outpvec;\cminpvec\given\chvec)=\tslots\fslots\,\mi(\tfoutpvec;\cmtfinpvec
\given\tfchmatw)$. The second term on the~RHS of \fref{eq:lb-mi-decomposition}
can be evaluated as
\be
\bs
	\mi(\outpvec;\chvec\given\cminpvec) & =\sum_{\rxindex=0}^{\numrx-1}
		\Exop\lefto[\logdet{\imat_{\tslots\fslots} + \rxevp\cminpmat
		(\txcorrspec\kron\schcorrmat)\herm{\cminpmat}}\right] \\
	&\stackrel{(a)}{\le}\sumz{\rxindex}{\numrx}\logdet{\imat_{\altnumtx\tslots
		\fslots} + \rxevp\Exop[\herm{\cminpmat}	\cminpmat](\txcorrspec
		\kron\schcorrmat)} \\
	&\stackrel{(b)}{=}\sumz{\txindex}{\altnumtx}\sumz{\rxindex}{\numrx}
		\logdet{\imat_{\tslots \fslots} + \txev_{\txindex}\rxevp
		\frac{\Pave\tstep}{\altnumtx\fslots}\schcorrmat}
\es\notag
\ee
where~(a) follows from Jensen's inequality because the log-determinant
expression is concave in~$\herm{\cminpmat}\cminpmat$~\cite{diggavi01-11a},
and~(b) follows because the $\{ \cminp_{\txindex}[\dtdf] \}$ are~\iid and have
zero mean and
constant modulus $\abs{\cminp_{\txindex}[\dtdf]}^{2}=\Pave\tstep/
(\altnumtx\fslots)$. We combine the two terms, set~$\bandwidth=\fslots
\fstep$, divide by~$\tslots\tstep$, and evaluate the limit for~$~\tslots\to
\infty$ by means of~\cite[Theorem~3.4]{miranda00-02a}, a generalization of
Szeg\"o's theorem for multilevel Toeplitz matrices. The resulting lower bound
can then be improved upon via time sharing: Let~$1\leq\tsparam\leq\papr$. We
transmit~$\sqrt{\tsparam}\cminpvec$ during a fraction~$1/\tsparam$ of the
transmission time and let the transmitter be silent otherwise.
\end{IEEEproof}

%--
\subsubsection*{Wideband Approximation of the Lower Bound}

For large enough bandwidth, and hence large enough~\fslots, the lower bound in
\fref{thm:lb} can be well approximated by an expression that is often much
easier to evaluate:
\begin{inparaenum}[(i)]
\item We replace the first term of~\LBonep by its Taylor
	series expansion up to first order, as given
	in~\cite[Theorem~3]{prelov04-08a}. This expansion requires the computation
	of the expectation of the trace of several terms that involve the channel
	matrix~$\sqm{\rxcorrspec}\tfchmatw\sqm{\txcorrspec}$. Lemmas~3 and~4
	in~\cite{lozano03-10a} provide the desired result.
\item An approximation of the second term results if we replace the
	$\fslots\times\fslots$ Toeplitz matrix~\chspecmatp by a circulant matrix
	that is, in~\fslots, asymptotically equivalent to~\chspecmatp~\cite{durisi08a}.
\end{inparaenum}
The resulting wideband approximation of~\LBonep then reads
\begin{multline}
\label{eq:lb-approx}
	\LBonep\approx\LBapproxp=\max_{1\le\tsparam\le\papr}\Biggl\{\frac{\numrx
		\Pave}{\altnumtx} \sumz{\txindex}{\altnumtx}\txev_{\txindex} \\
	- \tsparam\Pave^{2}\frac{\tfstep}{\bandwidth}\frac{\left(\sumz{\txindex}
		{\altnumtx}\txev_{\txindex}\right)^{2} \sumz{\rxindex}{\numrx}\rxevp^{2}
		+ \numrx^{2} \sumz{\txindex}{\altnumtx}\txev_{\txindex}^{2}}
		{2\altnumtx^{2}}\\
	-\frac{\bandwidth}{\tsparam}\sumz{\txindex}{\altnumtx}\sumz{\rxindex}
		{\numrx}\spreadint{\log\left(1+\txev_{\txindex}\rxevp
		\frac{\tsparam\Pave}{\altnumtx\bandwidth}\scfp\right)\!}\Biggr\}.
\end{multline}
This approximation is exact for~$\bandwidth\to\infty$~\cite{durisi08a}.

%------------------------------
\subsection{Numerical Examples}
\label{sec:numerical}

For a $3\times 3$ MIMO system, we show in this section plots of the upper bound~\UBonep
of~\fref{thm:ubTFpeak}, and---for~\altnumtx between~1 and~3---plots of the lower
bound~\LBonep of~\fref{thm:lb} and of the corresponding approximation~\LBapproxp in~\eqref{eq:lb-approx}.
The large-bandwidth behavior of the bounds will be substantiated in~\fref{sec:wideband-asymptotes}.

%--
\subsubsection*{Numerical Evaluation of the Lower Bound}

While the upper bound~\UBonep for~$\avpopt=\txevm$ can be efficiently evaluated,
direct numerical evaluation of the lower bound~\LBonep is difficult for
large~\fslots. First, it is necessary to numerically compute the mutual
information $\mi(\tfoutpvec;\sqrt{\tsparam}\cmtfinpvec\given\tfchmatw)$ for
constant modulus inputs; second, the eigenvalues of the~$\fslots \times \fslots$
matrix~\chspecmatp are required for the evaluation of the penalty term
in~\eqref{eq:lb}. While efficient numerical algorithms exist to solve the first
task~\cite{he05-05a}, the second task is  often challenging, especially
if~\fslots is large. In~\cite{durisi08a}, we present upper and lower bounds on
the penalty term in~\eqref{eq:lb} that are more amenable to numerical
evaluation. For the set of parameters considered in the next subsection, these
bounds are tight and allow to fully characterize~\LBonep numerically.

%--
\subsubsection*{Parameter Settings}

All plots are for a receive power normalized with respect to the noise spectral
density of~$\Pave/(1\,\mathrm{W}\!/\mathrm{Hz})=1.26\cdot10^{8}\,\inv{\mathrm{s}}\!$. This parameter value corresponds, for example, to a
transmit power of~$0.5\mW$, a thermal noise level at the receiver
of~$-174\dBm/\mathrm{Hz}$, free-space path loss over a distance of~$10\m$, and a
rather conservative receiver noise figure of~$20\dB$. Furthermore, we assume
that the scattering function is brick shaped with~$\maxDelay=5\mus$,
$\maxDoppler=50\Hz$, and corresponding spread~$\spread=10^{-3}$. Finally, we
set~$\papr=1$. For this set of parameter values, we analyze three different
scenarios: a spatially uncorrelated channel, spatial correlation at the receiver
only, and spatial correlation at the transmitter only.
%--
\subsubsection{Spatially Uncorrelated Channel}
\begin{figure}
\centering
	\includegraphics[width=\figwidth]{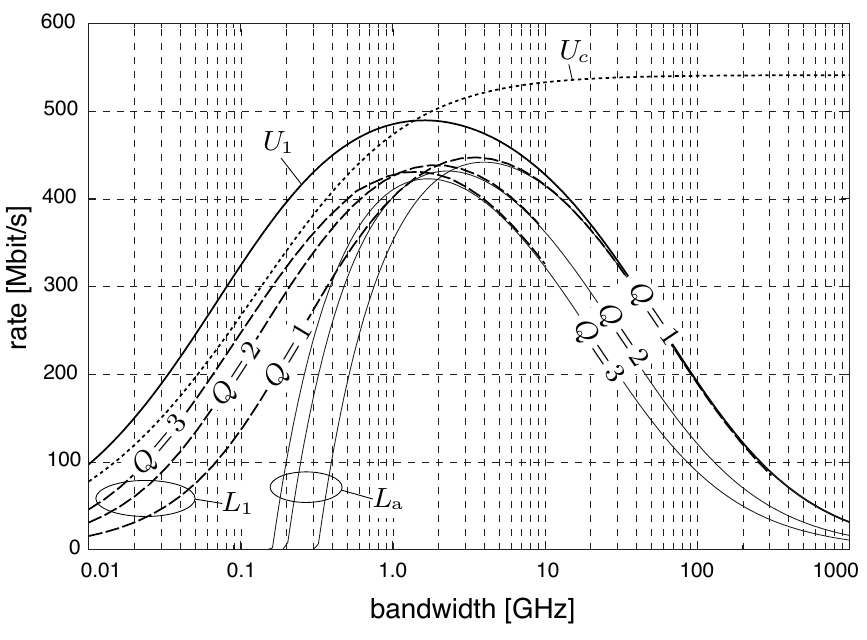}
  	\caption{Upper and lower bounds on the capacity of a spatially uncorrelated
		underspread~WSSUS channel with~$\txcorrspec=\rxcorrspec=\imat_{3}$,
		$\numtx=\numrx=3$, $\papr=1$, and~$\spread=10^{-3}$.}
	\label{fig:spatially-white}
\end{figure}

\fref{fig:spatially-white} shows the upper bound~\UBonep and---for~\altnumtx
between~1 and~3---the lower bound~\LBonep and the corresponding
approximation~\LBapproxp for the spatially uncorrelated case~$\txcorrspec=
\rxcorrspec=\imat_{3}$. For comparison, we also plot a standard capacity upper
bound~\UBcp obtained for the coherent setting and with input subject to an
average-power constraint only. We can observe that~\UBcp is tighter than~\UBonep
for small bandwidth; this holds true in general as for
small~\bandwidth the penalty term in~\eqref{eq:ubone} can be neglected
and~\UBonep in the spatially uncorrelated case reduces to
\bas
	\UBonep\approx  \frac{\numrx\bandwidth}{\tfstep}\log\Bigl(1+\frac{\Pave\tfstep}{\bandwidth}\Bigr)
\eas
which is the Jensen upper bound on the capacity~\UBcp in the coherent setting.
For small and medium bandwidth, the lower bound~\LBonep increases with~\altnumtx
and comes surprisingly close to the coherent capacity upper bound~\UBcp for~$\altnumtx=3$.

As can be expected in the light of e.g.,~\cite{medard02-04a,subramanian02-04a},
when bandwidth increases above a certain {\em critical bandwidth}, both~\UBonep
and~\LBonep start to decrease; in this regime, the rate gain resulting from the
additional degrees of freedom is offset by the resources required to resolve
channel uncertainty. The same argument seems to hold in the wideband regime for the degrees of freedom
provided by multiple transmit antennas: \UBonep appears to match \LBonep
for~$\altnumtx=1$; hence, using a single transmit antenna seems optimal in the
wideband regime.

%--
\subsubsection{Impact of Receive Correlation}
\begin{figure}
\centering
	\includegraphics[width=\figwidth]{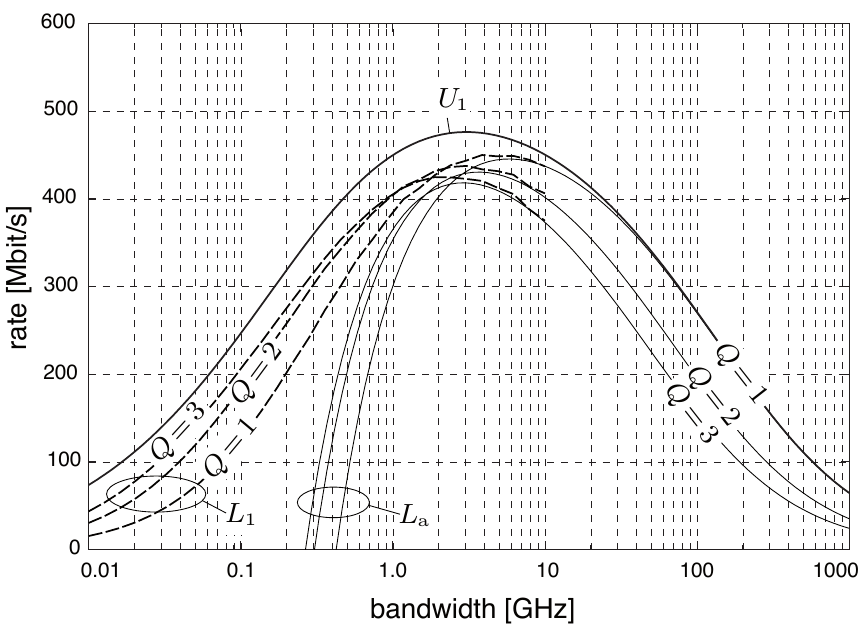}
  	\caption{Upper and lower bounds on the capacity of an underspread~WSSUS
  		channel that is spatially uncorrelated at the transmitter, $\txcorrspec=
		\imat_{3}$, but correlated at the receiver with $\rxcorrspec=
		\dg{\tp{[2.6\:\, 0.3\:\, 0.1]}}$; $\numtx=\numrx=3$, $\papr=1$,
		and~$\spread=10^{-3}$.}
	\label{fig:rx-correlated}
\end{figure}

\fref{fig:rx-correlated} shows the same bounds as before, but evaluated with
spatial correlation~$\rxcorrspec=\dg{\tp{[2.6\:\, 0.3\:\, 0.1]}}$ at the
receiver and a spatially uncorrelated channel at the transmitter,
i.e.,~$\txcorrspec=\imat_{3}$. The curves in \fref{fig:rx-correlated} are very
similar to the ones shown in \fref{fig:spatially-white} for the spatially
uncorrelated case, yet they are shifted towards higher bandwidth while the maximum rate is lower. Hence, at least for the example at hand, receive
correlation decreases capacity at small bandwidth but it is beneficial at large bandwidth.

%--
\subsubsection{Impact of Transmit Correlation}

\begin{figure}
\centering
	\includegraphics[width=\figwidth]{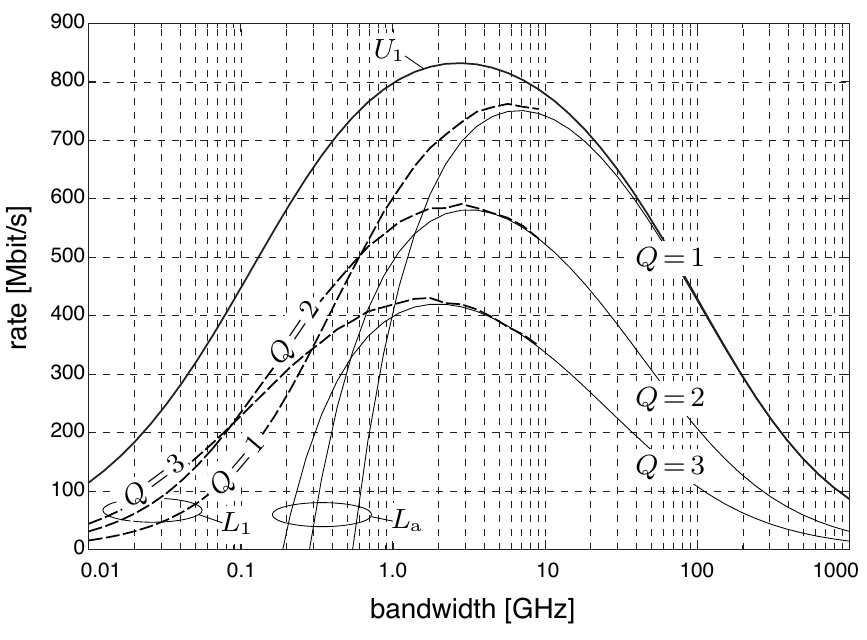}
  	\caption{Upper and lower bounds on the capacity of an underspread~WSSUS
  		channel that is correlated at the transmitter with $\txcorrspec =
		\dg{\tp{[1.7\:\, 1.0\:\,  0.3]}}$ and spatially uncorrelated at the
		receiver, $\rxcorrspec=\imat_{3}$; $\numtx=\numrx=3$, $\papr=1$,
		and~$\spread=10^{-3}$.}
	\label{fig:tx-correlated}
\end{figure}

We evaluate the same bounds once more, but this time for spatial
correlation~$\txcorrspec = \dg{\tp{[1.7\:\, 1.0\:\,  0.3]}}$ at the transmitter
and a spatially uncorrelated channel at the receiver, i.e., $\rxcorrspec=\imat_{3}$. The corresponding curves are shown in \fref{fig:tx-correlated}. Here, transmit correlation increases the capacity at large bandwidth,
while its impact at small bandwidth is more difficult to judge because the
distance between upper and lower bound increases compared to the spatially
uncorrelated case.

All three figures show that for large bandwidth the approximation~\LBapproxp
of~\LBonep is quite accurate. An observation of significant practical importance
is that the bounds~\UBonep and~\LBonep are quite flat over a large range of
bandwidth around their maxima. Further numerical results point at the following:
\begin{inparaenum}[(i)]
\item for smaller values of the channel spread~\spread, these maxima broaden and
	extend towards higher bandwidth; 
\item an increase in~\papr increases the gap between upper and lower bounds.
\end{inparaenum}

%%%%%%%%%%%%%%%%%%%%%%%%%%%%%
\section{The Wideband Regime}
\label{sec:wideband-asymptotes}

The numerical results in \fref{sec:numerical} suggest that in the wideband regime
\begin{inparaenum} [(i)] 
\item using a single transmit antenna is optimal when the channel is
	spatially uncorrelated at the transmitter side;
\item it is optimal to signal over the maximum transmit eigenmode if transmit
	correlation is present;
\item  both transmit and receive correlation are beneficial.
\end{inparaenum} 
To substantiate these observations, we compute the first-order
Taylor series expansion of~$\capacity(\bandwidth)$ around~$1/\bandwidth=0$. 

\begin{thm}
\label{thm:taylor}
Define
\ba
\label{eq:peakiness}
	\peakiness \define \spreadint{\scfpsq},
		\qquad\text{and}\qquad
	\rxevsqsum \define\sumz{\rxindex}{\numrx}\rxevp^{2}.
\ea
Then, for~$\papr>2\tfstep/\peakiness$, the
capacity~\fref{eq:channel-capacity} of the underspread WSSUS~MIMO channel
in \fref{sec:mat-vec-notation} under the power constraints in \fref{sec:power-constraints} has the following first-order Taylor series
expansion around~$1/\bandwidth =0$:
\begin{subequations}
\label{eq:taylor}
\ba
	&\capacity(\bandwidth)= \frac{\taylorone}{\bandwidth}
		+\osmall{\frac{1}{\bandwidth}} \label{eq:cap-taylor} \\
		\intertext{where}
	&\taylorone= \rxevsqsum\,\frac{(\txevm\Pave)^{2} }{2}\left(\papr\peakiness	
		-\tfstep\right)\label{eq:taylor-coeff}.
\ea
\end{subequations}
\end{thm}

%------------------------------------
\begin{IEEEproof}
The proof is a generalization of a similar proof for SISO channels
in~\cite[Appendices~E and~G]{durisi08a}; therefore, we only sketch the main
steps.

First, we expand the upper bound in \fref{thm:ubTFpeak} into a Taylor series. If
the channel is highly underspread, the sufficient
condition~\fref{eq:avpopt-conditions} for~$\avpopt=\txevm$ to achieve the
supremum in~\fref{eq:ubTFpeak-core} is valid for large enough bandwidth and
hence for~$\bandwidth\to\infty$. Therefore, we only need to expand~\UBonep
for~$\avpopt=\txevm$. A more refined analysis
in~\cite[Appendix~E]{durisi08a} shows that the supremum
in~\eqref{eq:ubTFpeak-core} is achieved for~$\avpopt=\txevm$ in the
large-bandwidth regime if and only if~$\papr>2\tfstep/\peakiness$, a condition
less restrictive than~\fref{eq:avpopt-spread}.

It follows from~\cite[Appendix~F]{durisi08a} that a Taylor series expansion of
the lower bound~\LBonep in \fref{thm:lb} does not match the corresponding
expansion of~\UBonep up to first order, so that we need to devise an
alternative, asymptotically tight, lower bound. We observed
in~\fref{sec:numerical} that signaling over a single transmit eigenmode seems to
be optimal for large bandwidth; hence, it is sensible to base the asymptotic
lower bound on a signaling scheme that uses only the strongest transmit eigenmode
in each slot. In one channel use, we thus transmit\footnote{Differently from the
coherent setting~\cite[Proposition~3]{tulino05-07a}, the multiplicity of the
largest eigenvalue of~\txcorrmat is immaterial. If this multiplicity is larger
than one, we choose to transmit along the eigenvector corresponding to index~$\txindex=0$ merely for
notational simplicity.}~$\inpvec=\tp{[\tp{\inpvec}_{0}\,\tp{\veczero}_{\tfslots}
\, \cdots\,\tp{\veczero}_{\tfslots}]}$, where~$\inpvec_{0}$ stands for the input
vector transmitted on the strongest eigenmode. Such a signaling scheme, often
referred to as {\em rank-one statistical beamforming}, transmits over all
available antennas in general; only if the channel is spatially uncorrelated at the
transmitter can antennas be physically switched off. With rank-one statistical
beamforming, the spatially decorrelated MIMO~channel with
IO~relation~\fref{eq:iorel-mimo} simplifies to a single-input
multiple-output~(SIMO) channel, the IO~relation of which can be conveniently expressed as
\bas
	\tilde{\outpvec} = \tilde{\chvec}\had\tilde{\inpvec}+\tilde{\wgnvec}
\eas
where~$\tilde{\wgnvec}$ is an $\numrx\tfslots$-dimensional JPG  vector with \iid
entries of zero mean and unit variance, the input vector $\tilde{\inpvec}\define
\tp{[\tp{\inpvec}_{0}\, \cdots\,\tp{\inpvec}_{0}]}$ contains~\numrx copies
of~$\inpvec_{0}$, and the stacked effective SIMO~channel vector
is~$\tilde{\chvec}\define\tp{[\tp{\chvec}_{0,0}\, \tp{\chvec}_{1,0}\, \cdots\,
\tp{\chvec}_{\numrx-1,0}]}$ with correlation matrix~$\covmat{\tilde{\chvec}}=\Ex{}{\tilde{\chvec}\herm{\tilde{\chvec}}}=\txevm\rxcorrspec\kron\schcorrmat$.
The desired asymptotic lower bound now follows directly from the derivation of
the asymptotic lower bound for a time-frequency selective SISO~channel
in~\cite[Appendix~G]{durisi08a}. In particular, we choose~$\inpvec_{0}$ to be
the product of a vector with \iid  zero mean constant modulus entries  and a nonnegative binary random variable with on-off distribution. Similar
signaling schemes were already used in~\cite{sethuraman08a} to prove asymptotic
capacity results for frequency-flat, time-selective channels. As the first-order
Taylor expansion of the resulting lower bound matches the first-order Taylor
expansion of~\UBonep in~\fref{eq:cap-taylor}, \fref{thm:taylor} follows.
\end{IEEEproof}

%--
\subsubsection*{Spatial Correlation and Number of Antennas}
Rank-one statistical beamforming along any eigenvector of~\txcorrmat associated
with~\txevm is optimal to attain the wideband asymptotes of~\fref{thm:taylor}.
For channels that are spatially uncorrelated at the transmitter, this result
implies that using only one transmit antenna is optimal, as previously shown
in~\cite{sethuraman08a} for the frequency-flat time-selective  case.
To further assess the impact of correlation on capacity, we
follow~\cite{chuah02-03a,jafar05-05a,jorswieck06-05a} and define a  partial
ordering of correlation matrices through {\em majorization}~\cite{marshall79a}.
We say that a correlation matrix~\matK entails more correlation than a
correlation matrix~\matC if the vector of eigenvalues~$\evvec(\matK)$
majorizes~$\evvec(\matC)$. To assess the impact of spatial correlation on
capacity, we further need the following definition~\cite{marshall79a}: a scalar
function~$\phi(\vecz)$ of a vector~\vecz is {\em Schur concave}
if~$\phi(\vecz)\leq \phi(\vecq)$ whenever~\vecz majorizes~\vecq.

In the {\em coherent setting}, capacity is Schur concave in~$\evvec(\rxcorrmat)
$, the eigenvalue vector of the receive correlation matrix while, for
sufficiently large bandwidth, it is Schur convex in~$\evvec(\txcorrmat)$, the
eigenvalue vector of the transmit correlation
matrix~\cite{tulino05-07a,jorswieck06-05a}. Hence, in the coherent setting
receive correlation is detrimental at any bandwidth while transmit correlation
is beneficial at large bandwidth. The intuition is that transmit correlation
allows to focus the transmit power into the maximum transmit eigenmode, and the
corresponding power gain offsets the reduction in effective transmit
signal space dimensions in the power-limited regime, i.e., at large bandwidth.
On the other hand, receive correlation is detrimental at any bandwidth because
it reduces the effective dimensionality of the receive signal space without any
power gain~\cite{lozano06a}. 

On the basis of~\fref{thm:taylor}, we conclude that the picture is fundamentally
different in the {\em noncoherent setting}. The coefficient~$\taylorone$
in~\eqref{eq:taylor} is a Schur-convex function in both the eigenvalue
vector~$[\txev_{0}\,\txev_{1}\,\cdots\,\txev_{\numtx-1}]$ of the transmit
correlation matrix and the eigenvalue vector~$[\rxev_{0}\,\rxev_{1}\,\cdots\,
\rxev_{\numrx-1}]$ of the receive correlation matrix because~\txevm
and~\rxevsqsum are continuous convex functions of the corresponding eigenvalue
vectors~\cite{marshall65-08a}. Hence, both receive and transmit correlation are
beneficial for sufficiently large bandwidth. This observation agrees with the
results for memoryless and block-fading channels reported
in~\cite{jafar05-05a,zhang07-03a,srinivasan07-10a}. In the wideband regime,
while transmit correlation is beneficial in both the coherent and the
noncoherent setting because it allows for power focusing, receive correlation is
beneficial rather than detrimental in the noncoherent setting for the following
reason: for fixed~$\numtx$ and~\numrx, the rate gain obtained from additional
bandwidth is offset in the wideband regime by the corresponding increase in
channel uncertainty (see Figs.~\ref{fig:spatially-white},~\ref{fig:rx-correlated}, and~\ref{fig:tx-correlated}); yet, for fixed but large bandwidth,
channel uncertainty decreases in the presence of receive correlation so that
capacity increases.

%--
\subsubsection*{The Lower Bound~\LBonep in the Wideband Regime}
Since we know that the first-order Taylor expansions around $(1/\bandwidth)=0$ of the upper bound~\UBonep and
the lower bound~\LBonep do not match, it is surprising that the corresponding
curves seem to coincide in Figs.~\ref{fig:spatially-white},~\ref{fig:rx-correlated}, and~\ref{fig:tx-correlated} for large bandwidth. The reason is that, for typical values of~\tfstep and~\papr, the ratio
between the first-order coefficients in the Taylor expansions of~\LBonep and~\capacityp
approaches~1 as~\peakiness grows large and~$\altnumtx=1$. For example, the
ratio is~$0.998$ for the parameters used in the numerical evaluation in
Fig.~\ref{fig:spatially-white}, i.e., $\spread=10^{-3}$, $\papr=1$,
and~$\tfstep=1.25$.

%%%%%%%%%%%%%%%%%%%%%%%%%%%%%%%%
\section{Discussion and Outlook}
\label{sec:conclusions}

Capacity analysis in the noncoherent setting is frequently performed
asymptotically for either large or small SNR,~$\Pave/\bandwidth$. The
corresponding asymptotic results are often useful to obtain design insight, but
they may sometimes be misleading: capacity behavior is very sensitive to
specific details of the channel model used at high~SNR~\cite{lapidoth05-07a},
and any channel model eventually breaks down for large enough bandwidth and
correspondingly low~SNR. The capacity bounds in the present paper are useful for
a large range of bandwidth in between these two asymptotic cases (in addition,
they are tight in the wideband regime).

The discrete-time discrete-frequency channel model presented in
\ref{sec:discrete-approx} and \fref{sec:mimo-extension} is very general; at
the same time, the corresponding capacity bounds in \fref{sec:bounds} are
relatively simple for practically relevant values of~$\Pave/\bandwidth$ and for
realistic scattering functions. Furthermore, as  our discrete-time
discrete-frequency channel model is related to the continuous time WSSUS~channel
model~\fref{eq:iorel-ct}, results from real-world channel measurements can be
directly used to obtain capacity estimates. In particular, as the bounds hold
for both the regime where degrees of freedom increase capacity, as well as for
the regime where degrees of freedom are detrimental, they allow to numerically
determine a suitable combination of bandwidth and number of transmit antennas.

For large bandwidth, the bounds are very accurate---the upper bound~\UBonep
exhibits the correct asymptotic behavior for~$\bandwidth\to\infty$, as shown in
\fref{sec:wideband-asymptotes}. 
For small and medium bandwidth, the upper bound~\UBonep is not tight, and is indeed worse than the 
coherent capacity upper bound. The fact that our simple lower bound~\LBonep comes quite close to the
coherent upper bound~\UBcp in \fref{fig:spatially-white} seems to validate, at
least for the setting considered, the standard receiver design principle to
first estimate the channel and then use the resulting estimates as if they were
perfect. To verify this conjecture, though, it is necessary to show that the
combination of dedicated channel estimation and coherent signaling achieves
rates similar to those predicted by the lower bound~\LBonep.

The advent of ultrawideband~(UWB) communication systems spurred the current
interest in wireless communications over channels with very large bandwidth.
Current UWB~regulations impose a limit on the power spectral density of
the transmitted signal, so that the available average power increases with
increasing transmission bandwidth. In contrast, we keep the total average
transmit power fixed in the present paper; therefore, the results presented here do not directly apply to current UWB~regulations. Nonetheless, our bounds allow
to assess whether multiple antennas at the transmitter are beneficial for
UWB~systems.
The system parameters used to numerically evaluate the bounds in
\ref{sec:numerical} are compatible with a UWB~system that operates over a
bandwidth of~$7\GHz$ and transmits at~-41.3\,dBm/\,MHz. Even if our bounds are not
tight at~$7\GHz$ in this scenario, Figs.~\ref{fig:spatially-white},~\ref{fig:rx-correlated}, and~\ref{fig:tx-correlated} show that the maximum rate increase
that can be expected from the use of multiple antennas at the transmitter does
not exceed~7\%. For channels with smaller spreads than the one
in~\fref{sec:numerical}, the possible rate increase is even smaller.

%%%%%%%%%%%
%%%%%%%%%%%
\appendices

%%%%%%%%%%%%%%%%%%%%%%%%%%%%%%%%%%
\section{A Determinant Inequality}
\label{app:det-inequality}

\begin{lem}
\label{lem:det-inequality}
Let~\matA and~\matB be two $\di\times\di$ nonnegative definite Hermitian
matrices. Then,
\bas
	\det(\imat_{\di} + \matA\had\matB)\ge\det\bigl(\imat_{\di} + (\imat_{\di}
		\had\matA)\matB\bigr).
\eas
\end{lem}

\begin{IEEEproof}
Assume for now that~\matA does not have zeros on its main diagonal and define
$\oppinvmat\define\inv{(\imat_{\di}\had\matA)}$. Then,
\be
\bs
	\det(\imat_{\di} + \matA\had\matB) &= \det\bigl(\matA \had (\oppinvmat+
		\matB) \bigr)\\
		&\stackrel{(a)}{\ge}\det(\imat_{\di}\had\matA)\det(\oppinvmat + \matB)\\
		&=\det\bigl((\imat_{\di}\had\matA)\oppinvmat + (\imat_{\di}\had\matA)
			\matB\bigr)\\
		&=\det\bigl(\imat_{\di} + (\imat_{\di}\had\matA)\matB\bigr)
\es
\label{eq:oppenheim-inequality-chain}
\ee
where~(a) is a direct consequence of Oppenheim's
inequality~\cite[Theorem~7.8.6]{horn85a}. To conclude the proof, we remove the
restriction that~\matA has only nonzero diagonal entries. Because~\matA is
nonnegative definite, its $\idx$th row and its $\idx$th column are zero
if~$[\matA]_{\idx\idx}=0$~\cite[Section 7.1]{horn85a}, so that, by the definition of the
Hadamard product, the $\idx$th row and the $\idx$th column of~$\matA\had\matB$
are zero as well. Let~\setI be the set that contains all indices~\idx for
which~$[\matA]_{\idx\idx}=0$, assume without loss of generality that there
are~\setsize such indices, and let~$\matA_{\setI}$ and~$\matB_{\setI}$ denote
the submatrices of~\matA and~\matB, respectively, with all rows and columns
corresponding to~\setI removed. An expansion by minors of~$\det(\imat_{\di} +
\matA\had\matB)$ now shows that
\ba
\label{eq:submatrix}
	\det(\imat_{\di} + \matA\had\matB) = \det(\imat_{\setsize} + \matA_{\setI}
		\had \matB_{\setI}).
\ea
Hence, it suffices to apply the inequality~\fref{eq:oppenheim-inequality-chain}
to the~RHS of~\fref{eq:submatrix}.
\end{IEEEproof}

%%%%%%%%%%%%%%%%%%%%%%%%%%%%%%%%%%%%%%%%%
\section{Optimization of the Upper Bound}
\label{app:avpopt}

The expression to be maximized in~\fref{eq:ubTFpeak-core},
\bas
	\ftop=\sumz{\rxindex}{\numrx}
		\Biggl(\frac{\bandwidth}{\tfstep}\log\Bigl(1+\avpopt
		\rxevp\frac{\Pave\tfstep}{\bandwidth}\Bigr)-\avpopt\ptp\Biggr)
\eas
where~\ptp is given in~\fref{eq:ubTFpeak-pt}, is concave in~\avpopt.
Hence,  the optimizing parameter~\avpopt is unique. Furthermore, the
following two properties hold: \begin{inparaenum}[(i)]
\item $\ftop=0$ for $\avpopt=0$.
\item As, by Jensen's inequality and because~$\log(1+x)\leq x$
	\be
	\label{eq:ub on penalty}
	\bs
    		\ptp&\leq  \frac{\bandwidth\spread}{\txevm\papr}\log\Bigl(1+
			\frac{\txevm\rxevp\papr\Pave}{\bandwidth\spread}\spreadint{\scfp}
			\Bigr)\\
		 &=	\frac{\bandwidth\spread}{\txevm\papr}\log\Bigl(1+
			\frac{\txevm\rxevp\papr\Pave}{\bandwidth\spread}\Bigr)\leq \rxevp
			\Pave,
	\es
	\ee
\end{inparaenum}
the first derivative of~\ftop,
\ba
\label{eq:avpopt-derivative}
	\ftoderp=\sumz{\rxindex}{\numrx}\Bigl(\frac{\rxevp\Pave}{1+\avpopt
		\rxevp\Pave\tfstep/\bandwidth}-\ptp\Bigr)
\ea
is nonnegative at~$\avpopt=0$.

From property~(i) and~(ii), and from the concavity of~\ftop, it follows that
the supremum in~\eqref{eq:ubTFpeak-core} is achieved for~$\avpopt=\txevm$
if and only if the zero of~\eqref{eq:avpopt-derivative} occurs at a point
larger or equal to~\txevm, or, equivalently, if and only
if~\eqref{eq:avpopt-derivative} is positive for~$\avpopt \in [0,\txevm)$.
Identification of this zero-crossing is difficult for~$\rank(\rxcorrmat)>1$, but
we can obtain a sufficient condition for the supremum to be achieved
for~$\avpopt=\txevm$ as follows:
\begin{compactitem}
\item The first derivative~\fref{eq:avpopt-derivative} will certainly be
	positive if all terms in the sum are positive.
\item As for all~\avpopt in the set~$[0,\txevm)$ the inequality
	\bas
		\frac{\rxevp\Pave}{1+\avpopt\rxevp\Pave\tfstep/
			\bandwidth} \ge \frac{\rxevp\Pave}{1+\txevm\rxevp
			\Pave\tfstep/\bandwidth}
	\eas
	holds, it follows from Jensen's inequality applied to~\ptp as
	in~\eqref{eq:ub on penalty} that a sufficient condition for the $\rxindex$th
	term in~\fref{eq:avpopt-derivative} to be positive in~$[0,\txevm)$ is
	\bas
		\frac{\rxevp\Pave}{1+\txevm\rxevp\Pave\tfstep/
		\bandwidth}> \frac{\bandwidth\spread}{\txevm\papr}\log\Bigl(1+
			\frac{\txevm\rxevp\papr\Pave}{\bandwidth\spread}\Bigr).
	\eas
\item This condition is very similar to one analyzed
	in~\cite[Appendix~C]{durisi08a}, and steps identical to the ones detailed
	in~\cite[Appendix~C]{durisi08a} finally lead to~\fref{eq:avpopt-conditions}.
\end{compactitem}

% trigger a \newpage just before the given reference
% number - used to balance the columns on the last page
% adjust value as needed - may need to be readjusted if
% the document is modified later
%\IEEEtriggeratref{26}
% The "triggered" command can be changed if desired:
%\IEEEtriggercmd{\enlargethispage{-5in}}

\bibliographystyle{IEEEtran}
\bibliography{IEEEabrv,publishers,confs-jrnls,ulibib}

\end{document}